\documentclass[a4paper,11pt]{article}
\pdfoutput=1 

\usepackage{jcappub} 

\usepackage[T1]{fontenc} 
\usepackage[utf8]{inputenc}
\usepackage{amsmath}
\usepackage{xcolor}

\newcommand*\diff{\mathop{}\!\mathrm{d}}
\newcommand{\sun}{\odot}

\title{Hierarchal formation and growth of galaxies through dynamical friction}

\author[a]{F. Huško}

\affiliation[a]{Department of Physics, Faculty of Science, University of Zagreb, Bijenička c. 32, 10000 Zagreb, Croatia}

\emailAdd{fhusko@dominis.phy.hr}

\abstract{We study the infall of a subhalo in its parent halo due to dynamical friction. Using expected mass and spatial distributions of subhaloes, in unison with the derived infall time-scale, we calculate the expected merger stellar mass growth rate (SMGR) of the central galaxy in two different models (instantaneous and continuous merging). We find that our continuous merging SMGR (which corresponds to smooth accretion from satellites) agrees with the results of the Millennium-II simulation, predicting low growth rates for low stellar masses ($10^{8}$ M$_{\sun}-10^{10}$ M$_{\sun}$) and accelerating growth for high masses ($10^{10}$ M$_{\sun}-10^{12}$ M$_{\sun}$). Using the derived formulas we study the relative contributions of minor and major mergers to the total merger SMGR at various redshifts. Minor mergers contribute $80\%-90\%$ in the low mass range for $z<4$, with major mergers becoming the dominant merger type (80$\%-95\%$) at relatively high masses ($4\times 10^{10}$ M$_{\sun}$) for $z>1$. This suggests that most merger mass growth is due to minor mergers. We derive an SMGR formula for the very early universe ($z>4$) and use this to show that hierarchial merging of protogalaxies can lead to the modern population of galaxies. Our overall analysis suggests that dynamical friction is the primary cause of galaxy mergers.}

\begin{document}
\maketitle
\flushbottom

\section{Introduction}
\label{sec:intro}

As shown by Chandrasekhar \cite{Chandrasekhar1943}, any massive body moving through space is slowed down by an effective frictional force, provided the background density is non-vanishing. This gravitational effect, usually referred to as dynamical friction, has since been used extensively as an explanatory mechanism for various physical phenomena. For example, it has a significant role in planetary formation \cite{O'Brien2006}. In the galactic and intergalactic regimes, dynamical friction is responsible for numerous processes, such as gas heating in galaxy clusters \cite{El-Zant2004} and accretion of globular clusters onto parent galaxies \cite{Lotz2001, Capuzzo-Dolcetta2005}. The same mechanism causes accretion of satellite galaxies in merger events, and thus dynamical friction is an integral component of any model that is to reproduce observables related to mergers, such as galaxy growth rates.

The principal way of quantifying the effect of dynamical friction in this problem is through the calculation of infall time-scales of a dark matter subhalo and its associated (satellite) galaxy in the halo of a parent galaxy. This is usually done by assuming a circular orbit and using Chandrasekhar's formula for deceleration due to dynamical friction. This should then be generalized to non-circular orbits \cite{Taffoni2003, BK2008}. Through N-body simulations authors in \cite{BK2008} found that non-circular infall time-scales (when compared with circular orbits) can differ up to a factor of 3 for the smallest relevant merger ratios ($M_{\textnormal{sat}}/M_{\textnormal{host}}=0.01$, where $M$ refers to (sub)halo mass), while stellar mass growth due to mergers can differ up to 40\%. Infall time-scales can also change as a result of tidal stripping of subhaloes by hosts - this changes their mass, thus decreasing their susceptibility to dynamical friction (due to the linear dependence of $a_{\textnormal{df}}$ on subhalo mass).

Once the correct infall time-scale formula is determined, it is possible to calculate various quantities associated with mergers. The usual quantity of interest in studies of mergers is the merger rate, which describes merger frequency at a given galaxy mass \cite{LaceyCole1993, GuoWhite2008, Stewart2009, Lotz2011}. Another quantity of interest is the merger stellar mass growth rate $\diff M_*/\diff t$, which has been studied through simulations in recent years \cite{Moster2013, Rodriguez-Gomez2016}. To our knowledge, no efforts have been given to deriving this quantity (semi-)analytically. Since it is a significant factor in galaxy formation and evolution, alongside star formation and gas accretion, it is our aim to find the functional dependence of the merger stellar mass growth rate on various galactic and halo parameters.

In Section \ref{sec:section2} we begin by setting up our galaxy and halo model. In Section \ref{sec:section3} we calculate the infall time-scale due to dynamical friction for a subhalo of arbitrary mass and initial position in a halo of arbitrary properties. We generalize from circular to non-circular orbits. In Section \ref{sec:section4} we determine the merger mass growth rate of a galaxy by considering the derived infall time-scale formula, as well as expected mass and spatial distributions of satellites. We present two different models$-$one for instantaneous and one for continuous merging. In Section \ref{sec:section5} we evaluate the growth rates as functions of parent galaxy stellar mass $M_*$, and compare our findings with previous results both at $z=0$ and for small $z$. We analyse the impact of minor and major galaxies on the total merger mass growth rate as a function of stellar mass and redshift. Furthermore, we evaluate the merger rate $\diff N/\diff t$ as a function of galaxy and halo properties. We show that our model is consistent with the hierarchal picture of galaxy formation by finding the growth history of early galaxies. We offer approximate formulas for both the SMGR and the merger rate. In Section \ref{sec:section6} we summarize and conclude.

\section{Model parameters}
\label{sec:section2}

Before calculating the infall time-scale of a satellite subhalo, here we set up our model and connect the various parameters that will appear in later calculations. It is our aim to determine the environment of an infalling subhalo of mass $m$, given only the redshift in consideration and stellar mass $M_*$ of the central galaxy.

Many of the following relations are only mean trends and exhibit some scatter, which can in principle be a function of both halo/galaxy mass and redshift. Here we present only the mean trends, and explain why in Section \ref{sec:subsubsection5.7.1}. With this in mind, it should be noted that all results presented in this work are only predictions for a typical galaxy. While it is highly improbable for any one galaxy to fulfill all the trends presented here, it is not our aim to predict the growth rate of an arbitrary galaxy of mass $M_*$ and at redshift $z$, but rather to predict the growth rate of an ensamble of galaxies of such mass and at that redshift.

\subsection{Spatial density of dark matter}
\label{sec:subsection2.1}

We assume the dark matter (DM) distribution of a halo is given by the Navarro-Frenk-White (NFW) profile \cite{NFW1996},

\begin{equation}
    \rho(r)=\frac{\rho_0}{r/r_s(1+r/r_s)^2},
	\label{eq:NFWprofile}
\end{equation}
where $\rho_0$ and $r_s$ are the characteristic density and scale radius of the halo in question. A simple integration yields the enclosed DM mass within a radius $r$ in an NFW halo:
\begin{equation}
    M(r)=4\pi \rho_0 r_s^3 \bigg[\ln\Big(1+\frac{r}{r_s}\Big)-\frac{r}{r+r_s}\bigg].
	\label{eq:EnclosedMass}
\end{equation}

The total mass of the halo $M$ is defined as the mass enclosed within the radius $R_{\textnormal{200}}$ of the halo, such that the {\it average} enclosed density is 200 times greater than the critical density of the universe at the cosmic epoch of interest. In other words, 
\begin{equation}
    M=\frac{800\pi}{3}\rho_{\textnormal{crit}} R_{\textnormal{200}}^3=\frac{800\pi}{3}\rho_{\textnormal{crit}}(r_sc)^3,
	\label{eq:TotalMass}
\end{equation}
where $c$ is the halo concentration, equal to the halo radius (R$_\textnormal{200}$) expressed in units of the scale radius. Evaluating the enclosed mass (equation~\ref{eq:EnclosedMass}) at the virial radius gives, along with equation~(\ref{eq:TotalMass}), a system of two equations for $\rho_0$ and $r_s$, the two NFW parameters. Solving these yields the two parameters as functions of halo mass, halo concentration and critical density:
\begin{equation}
    \rho_0=\frac{200}{3}\rho_{\textnormal{crit}}\frac{c^3}{\ln(1+c)-c/(1+c)},
	\label{eq:Rho0}
\end{equation}
\begin{equation}
    r_s=\frac{1}{c}\bigg( \frac{3}{800\pi} \frac{M}{\rho_{\textnormal{crit}}} \bigg)^{1/3}.
	\label{eq:ScaleA}
\end{equation}

Halo mass and concentration at a given redshift are generally dependent. Here we use the relation from \cite{Munoz2011}:
\begin{equation}
    \log_{10} c=A(z)\log_{10}[M/(h^{-1}M_{\sun})]+B(z),
	\label{eq:Concentration(M)}
\end{equation}
where the functions $A(z)$ and $B(z)$ are given by
\begin{equation}
    A(z)=0.029z-0.097,
	\label{eq:A(z)}
\end{equation}
\begin{equation}
    B(z)=-\frac{110}{z+16.86}+\frac{2470}{(z+16.86)^2}.
	\label{eq:B(z)}
\end{equation}

We assume the same cosmological parameters as in \cite{Munoz2011}, namely: $h=0.72,$ $\Omega_{\textnormal{m}}=0.26$, $\Omega_\Lambda=0.74$. The critical density of the universe at redshift $z$ is given by
\begin{equation}
   \rho_{\textnormal{crit}}(z)=\frac{3H(z)^2}{8\pi G},
	\label{eq:B(z)}
\end{equation}
where the Hubble parameter is approximately $H(z)=H_0\sqrt{0.26(1+z)^3+0.74}$.

The critical density only depends on redshift, while the concentration depends on both halo mass and redshift through equation (\ref{eq:Concentration(M)}). Equations (\ref{eq:Rho0}) and (\ref{eq:ScaleA}) then determine the dependences of the characteristic density $\rho_0$ and scale radius $r_s$ on halo mass and redshift. In this way, given a halo mass $M$, the spatial density of dark matter is fully determined (up to a choice of redshift).

\subsection{Stellar mass content}
\label{sec:subsection2.2}

 In principle, haloes can be devoid of galaxies or host multiple `central' (non-satellite) galaxies. Here we restrict ourselves to considering the case of a primary halo hosting one central galaxy.

The expected stellar mass $M_*$ of a central galaxy residing in a halo of mass $M$ is given by the stellar-to-halo mass relation (SHMR). This can be derived from models of halo occupation distribution \cite{Yang2012}, through abundance matching techinques \cite{Moster2013}, or by directly fitting observational data \cite{Hansen2009}. The results from all of these methods are generally in good agreement. Here we choose the relation from \cite{Moster2013}, given by
\begin{equation}
   M_*(M)=2N_0 (z)M\bigg\{ \bigg[\frac{M}{M_0(z)}\bigg]^{-\beta(z)}+\bigg[\frac{M}{M_0(z)}\bigg]^{\gamma(z)} \bigg\}^{-1},
	\label{eq:SHMR}
\end{equation}
with the redshift dependence of the parameters $\{N_0(z)$, $M_0(z)$, $\beta(z)$, $\gamma(z)\}$ given in Tab.~\ref{tab:SHMRpar}.

In order to express various quantities (such as the characteristic density or scale radius of the halo) as functions of galaxy stellar mass instead of halo mass, we find the inverse relation to equation~(\ref{eq:SHMR}) through numerical means wherever needed. We denote this function as $M(M_*)$ in the rest of the text.

\begin{table}[tbp]
\centering
\begin{tabular}{|lr|c|}
\hline
$N_0(z)$& & $\hspace{2mm}0.035-0.025\frac{z}{1+z}$\\
\hline
$\log_{10} M_0(z)$& & $11.59+1.12\frac{z}{1+z}$\\
\hline
$\beta(z)$& & $1.38-0.83\frac{z}{1+z}$\\
\hline
$\gamma(z)$& & $0.61+0.33\frac{z}{1+z}$\\
\hline
\end{tabular}
\caption{\label{tab:SHMRpar} Redshift dependence of the SHMR parameters in equation~(\ref{eq:SHMR}).}
\end{table}

\subsection{Galaxy size}
\label{sec:subsection2.3}

In order to define the size of a galaxy we first consider its effective radius. This has been studied extensively using SDSS data \cite{Shen2003}. The authors in \cite{Shen2003} found that the effective radii of early-type (elliptical) and late-type (spiral) galaxies are generally different at a given stellar mass $M_*$. The relations in question are given by:
\begin{equation}
   R_{\textnormal{eff,E}}=R_{0,\textnormal{E}}\bigg(\frac{M_*}{\textnormal{M}_{\sun}}\bigg)^{0.56}\hspace{1mm} \textnormal{kpc},
	\label{eq:EffRadiusE}
\end{equation}
\begin{equation}
   R_{\textnormal{eff,L}}=R_{0,\textnormal{L}}\bigg(\frac{M_*}{\textnormal{M}_{\sun}}\bigg)^{0.14} \bigg(1+\frac{M_*}{M_1} \bigg)^{0.25} \hspace{1mm} \textnormal{kpc},
	\label{eq:EffRadiusE}
\end{equation}
with the constants \{$R_{0,\textnormal{E}}$, $R_{0,\textnormal{L}}$, M$_1$\} equal to \{$3.47\times 10^{-6}$, 0.1, $3.98\times 10^{10}\textnormal{M}_{\sun}$\}, respectively.

While the effective radius can be a sufficient measure of galaxy size in certain situations, we find $R_{90}$, the radius corresponding to 90\% of emitted luminosity, to be a more appropriate quantity in the context of mergers. The authors of \cite{Yamauchi2005} studied the concentration index $C=R_{\textnormal{eff}}/R_{90}$, and found that $C_E=0.29$ for galaxies that follow de Vaucouleurs' profile, while $C_L=0.44$ for disk galaxies. It should be noted that not all early-type galaxies follow de Vaucouleurs' profile. In this sense, taking the concentration index of any early-type galaxy to be 0.29 is a rather rough approximation. 

An additional reason why $R_{90}$ might be a better measure of galaxy size in this context (instead of the effective radius) is the existence of a truncation radius in disk galaxies. This truncation of exponential disks occurs at the mean value of $4R_d$ \cite{Kregel2002}, where $R_d$ is the scale length of the disk. Furthermore, there is a linear relationship between mean values (at a given stellar mass) of the scale length and effective radius \cite{Fathi2010}, $R_{\textnormal{eff}}=1.5R_d$. This means that the truncation ocurrs at $4R_d=8/3R_{\textnormal{eff}}$, with $8/3\approx2.33$, while $1/C_L\approx2.27$. Thus, the truncation radius is approximately equal to $R_{90}$ for disk galaxies.

\begin{figure}
\centering
	\includegraphics[width=0.6\columnwidth]{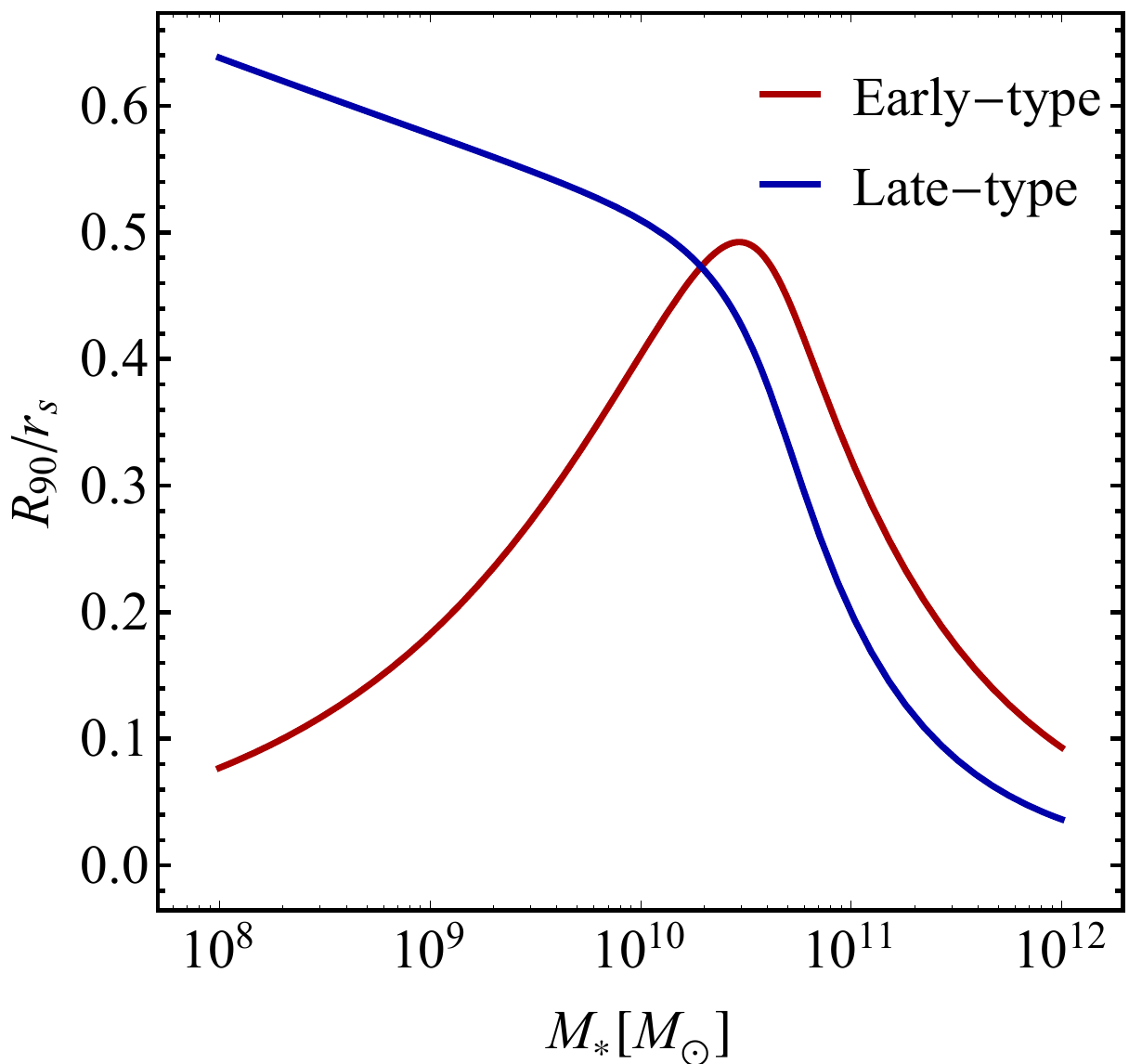}
    \caption{Galaxy-halo ratio, obtained by dividing the galaxy size by the dark matter scale radius, as a function of the central galaxy stellar mass $M_*$. The red and blue lines correspond to early and late-type galaxies, respectively.}
    \label{fig:fig1}
\end{figure}

With this in mind, we now define galaxy sizes as
\begin{equation}
   R_{\textnormal{E}}=R_{\textnormal{eff,E}}/C_E=3.45R_{\textnormal{eff,E}},
	\label{eq:GalSizeE}
\end{equation}
\begin{equation}
   R_{\textnormal{L}}=R_{\textnormal{eff,L}}/C_L=2.27R_{\textnormal{eff,L}},
	\label{eq:galSizeL}
\end{equation}
 for early-type and late-type galaxies, respectively. The relevant quantity in our calculations will be the ratio $R/r_s$ (galaxy-halo ratio, GH), where $r_s$ is the dark matter scale radius. In a sense, the GH ratio measures how deep inside the core of the halo the galaxy is embedded in since $r_s$ is the distance at which the density changes its behaviour from $\rho\propto r^{-1}$ to $\rho \propto r^{-3}$. The dependence of the GH ratio on stellar mass of the central galaxy is shown on Fig. \ref{fig:fig1}.  Late-type galaxies exhibit a slow decline in their GH ratio (65\% down to 50\%) in the low mass regime ($10^8\textnormal{M}_{\sun} - 10^{10}\textnormal{M}_{\sun}$), with a steep decline to 10\% occuring in the $10^{10}\textnormal{M}_{\sun} - 10^{11}\textnormal{M}_{\sun}$ interval. In the same range, early-type galaxies exhibit a maximum in their GH ratio$-$both very low and very high mass early-type galaxies are embedded deep in the cores of their parent haloes.

\section{Dynamical friction infall time-scale}
\label{sec:section3}

With the model set up, and the expected environment of a subhalo determined, we can now calculate the infall time-scale of a DM subhalo of mass $m$ from initial position $R_0$. We use the formula derived by Chandrasekhar \cite{Chandrasekhar1943} for deceleration due to a background density $\rho$ with a Maxwellian distribution of velocities (characterised by dispersion $\sigma_\rho$), given by
\begin{equation}
   \boldsymbol{a}_\textnormal{df}=-4\pi\ln \Lambda G^2 m\rho f(v_\rho, \sigma_\rho)\frac{\boldsymbol{v}}{v^3},
	\label{eq:Chandrasekhar}
\end{equation}
where $\ln \Lambda$ is the Coulomb logarithm (characteristic of dispersive interactions), $\boldsymbol{v}$ the velocity of the subhalo and $v_\rho$ the background particle velocity. The function $f$ takes the form
\begin{equation}
  f(v_\rho, \sigma_\rho)=\textnormal{erf}\Bigg(\frac{v_\rho}{\sqrt{2\sigma_\rho^2}} \Bigg)-\frac{2}{\sqrt{\pi}}\frac{v_\rho}{\sqrt{2\sigma_\rho^2}}\exp\Bigg(- \frac{v_\rho^2}{2\sigma_\rho^2} \Bigg).
	\label{eq:Funf}
\end{equation}

We assume dark matter to be the only relevant density component \cite{BK2008}; the density is then given by the NFW profile as determined in the previous section (equation \ref{eq:NFWprofile}). The circular velocity of DM particles is
\begin{equation}
   v_\rho=\sqrt{\frac{GM(r)}{r}}=\sqrt{4\pi G \rho_0 r_s^2}\sqrt{\frac{1}{x}\ln (1+x)-\frac{1}{1+x}},
	\label{eq:CircVel}
\end{equation}
where we have substituted $M(r)$ from equation (\ref{eq:EnclosedMass}), and with $x=r/r_s$. The dispersion $\sigma_\rho$ can be obtained by solving Jeans' equation. A fitting formula found in \cite{Zentner2003} has the form
\begin{equation}
   \sigma_\rho=v_\textnormal{max}\frac{1.44x^{0.354}}{1+1.176x^{0.725}},
	\label{eq:Dispersion}
\end{equation}
with $v_\textnormal{max}$ the maximal velocity of the NFW profile, achieved at $r\approx2.2r_s$. Substituting this value in equation (\ref{eq:CircVel}), we have $v_\textnormal{max}\approx 1.65\sqrt{G\rho_0 r_s^2}$.

We take the Coulomb logarithm $\ln \Lambda$ to be independent of position and only dependent on the mass ratio of subhalo to parent halo, $\ln \Lambda=\ln (1+M/m)$ \cite{Taffoni2003, Jiang2008}.

\subsection{Circular orbits}
\label{sec:subsection3.1}

We first restrict ourselves to considering circular orbits such that the velocity of the subhalo at position $r$ is the same as that of background particles; $v=v_\rho$. With this assumption, the deceleration given by equation (\ref{eq:Chandrasekhar}) becomes dependent only on position $r$.

We find the infall time-scale by considering the loss of angular momentum,
\begin{equation}
   \frac{\diff L}{\diff t}=I\alpha,
	\label{eq:AngMomentum}
\end{equation}
with $\alpha=a_\textnormal{df}/r$. Assuming a point-mass satellite, the moment of inertia is given by $I=mr^2$. Assuming no mass loss, the dissipation of angular momentum can be written out as
\begin{equation}
   \frac{\diff L}{\diff t}=m\frac{\diff}{\diff t}(rv)=m\bigg(\dot{r}v+r\frac{\diff v}{\diff t}\bigg)=m\dot{r}\bigg(v+r\frac{\diff v}{\diff r}\bigg).
	\label{eq:AngMomentum2}
\end{equation}
Combining the last expression with equation (\ref{eq:AngMomentum}), we find the infall velocity to be
\begin{equation}
   \dot{r}=\frac{ra_\textnormal{df}}{v+r\diff v/\diff r}.
	\label{eq:InfallVelocity}
\end{equation}
Substituting $a_\textnormal{df}$ and seperating, we solve for the infall time-scale:
\begin{equation}
    \tau=\frac{1}{4\pi \ln \Lambda m G^2}\int_R^{R_0}\frac{v^2(v+r\diff v/\diff r)}{r\rho f(v, \sigma)}\diff r,
	\label{eq:InfallTimescaleCirc}
\end{equation}
with $R_0$ the initial radius and $R$ the radius at which we consider the subhalo merged with the central galaxy. Substituting in the velocity and density, as well as choosing a dimensionless integration variable $x=r/r_s$, we have
\begin{equation}
    \tau=\frac{2\sqrt{\pi \rho_0 /G}r_s^3}{m \ln \Lambda}\int_{R/r_s}^{R_0/r_s}\frac{g^2(g+x\diff g/\diff x)}{(1+x)^2 f[g(x), \sigma(x)]}\diff x,
	\label{eq:InfallTimescaleCirc2}
\end{equation}
where $g(x)$ is given by
\begin{equation}
    g(x)=\sqrt{\frac{1}{x}\ln (1+x)-\frac{1}{1+x}}.
	\label{eq:g(x)}
\end{equation}
The rather complicated integrand can be approximated very closely by $1.35\sqrt{x}(1+x)^{3/4}$ for $x$ between 0.1 and 15. We expect a subhalo will spend most of its time in regions outside the core of the halo (sufficiently far from the galaxy), which means that values for very small $x$ are not relevant. The upper boundary, equal to 15, coincides with the maximal concentration of haloes in consideration, as shown in the previous section (we do not consider infall from radii outside the virial radius).

With the approximated integrand, the integral in equation (\ref{eq:InfallTimescaleCirc2}) can be solved explicitly in terms of a hypergeometric function. This result can further be approximated closely by $1.35\times0.71x^{2.1}$. For simplicity, we choose a sligthly different approximation, $1.35\times0.85x^2$. Finally, the infall time-scale of a subhalo on a circular orbit is found to be
\begin{equation}
    \tau_\textnormal{circ}=\frac{23}{10}\sqrt{\frac{\pi \rho_0}{G}}\frac{r_s}{m\ln \Lambda}R_0^2,
	\label{eq:InfallTimescaleCircResult}
\end{equation}
with the numerical factor approximately equal to 4.08. This result is achieved if the final radius $R$ is equal to zero. Even though subhaloes do not merge immediately upon reaching the central galaxy, but only once they lose all of their specific angular momentum, the time-scale of this process is consistent with that of the subhalo reaching the galactic center \cite{BK2008}. It should be noted that subhaloes effectively never reach the galactic center$-$they are tidally disrupted before this can occur. The effect on time-scales is minimal$-$since it scales quadratically with initial position, the subhalo will spend most of its time in the outer regions as long as its initial radius is taken to be outside the inner regions of its parent halo (and we expect most subhaloes to be outside the core).

Interestingly, the infall time-scale, as given by equation (\ref{eq:InfallTimescaleCircResult}), has the same dependence on initial position as that in an isothermal density profile. A similar result was found by \cite{Taffoni2003}, $\tau \propto x^{1.97}$. This is to be compared to our result $\tau \propto x^{2.1}$. The numerical factor $4.08$ agrees well with fits done on the assumption that the profile is isothermal, where the factor is equal to 4.12 \cite{LaceyCole1993}.

\subsection{Non-circular orbits}
\label{sec:subsection3.2}

The generalization of circular orbit infall time-scales to non-circular ones is usually done through an additional parameter: the orbital circularity $\epsilon=j/j_\textnormal{circ}(E)$, which measures the specific angular momentum of the actual orbit as compared to a circular orbit of equal energy. This parameter allows the time-scale to vary with changing angular momenta. We write the generalized infall time-scale as:
\begin{equation}
    \tau=h(\epsilon)\tau_\textnormal{circ}.
	\label{eq:NonCirc}
\end{equation}

The exact form of the function $h(\epsilon)$ is usually retrived from simulations. Results from different authors generally do not agree well. For example, the authors of \cite{BK2008} find $h$ to be an exponential function of $\epsilon$, while those of \cite{Cole2000} use $h(\epsilon)=\epsilon^{0.78}$. We find the best agreement with simulations (as offered through comparisons in Section \ref{sec:section5}) to be the result found in \cite{Jiang2008}, where the relation in question is
\begin{equation}
    \tau=(0.61\epsilon^{0.6}+0.39) \hspace{0.3mm}\tau_\textnormal{circ}.
	\label{eq:NonCircEpsAndEta}
\end{equation}

Since we are interested only in a statistical measure of stellar mass growth (across a large number of galaxies at a given stellar mass), it is sufficient to integrate equation (\ref{eq:NonCircEpsAndEta}) over the expected distribution of $\epsilon$, as given by cosmological simulations. We take the probability distribution $p(\epsilon)$ as in \cite{Jiang2008},
\begin{equation}
    p(\epsilon)=2.77\epsilon^{1.19}(1.55-\epsilon)^{2.99}.
	\label{eq:DistrEpsilon}
\end{equation}
Integrating equation (\ref{eq:NonCircEpsAndEta}), after a multiplication by the distribution $p(\epsilon)$, yields a numerical factor of 0.78. Thus, we finally have the {\it mean} (expected) infall time-scale of the subhalo in question:
\begin{equation}
    \tau \approx \frac{16}{5}\sqrt{\frac{\rho_0}{G}}\frac{r_s}{m\ln (1+M/m)}R_0^2.
	\label{eq:FinalTimescale}
\end{equation}

\begin{figure*}
	\includegraphics[width=\textwidth]{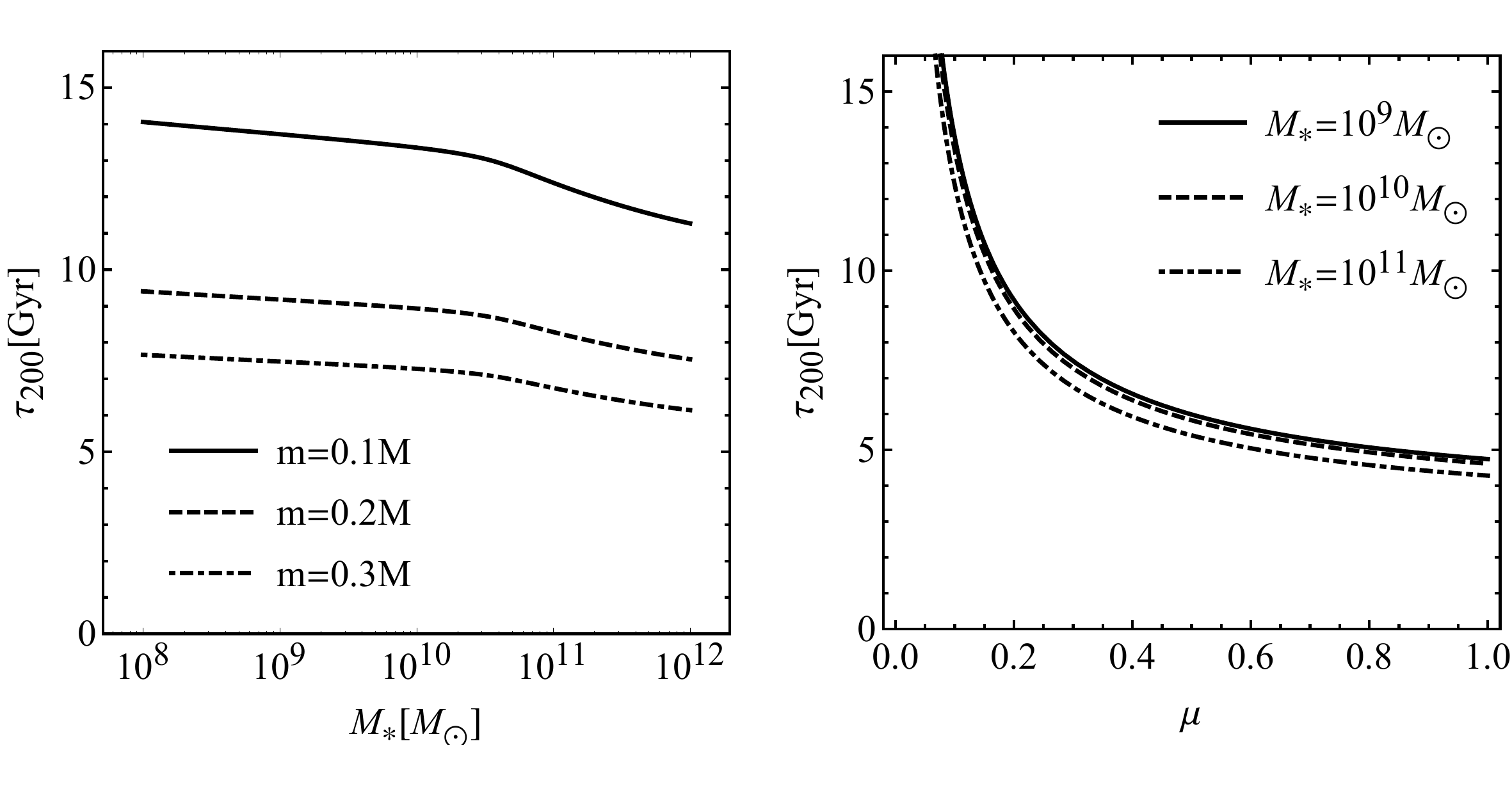}
    \caption{Infall time-scale from the edge of the host halo, given by equation (\ref{eq:FinalTimescale}) with the substitution $R_0=r_sc$, and parametrization $m=\mu M$. {\it Left}: Infall time-scale as a function of central galaxy stellar mass for three different subhalo-primary halo mass fractions $\mu$ (according to the legend). {\it Right}: Infall time-scale as a function of subhalo-primary halo mass fraction for three different central galaxy stellar masses (according to the legend).}
    \label{fig:fig2}
\end{figure*}

Fig. \ref{fig:fig2} shows the infall time-scale from the halo edge $R_0=R_\textnormal{200}=r_sc$, with $\rho_0$, $r_s$ and $c$ in equation (\ref{eq:FinalTimescale}) evaluated as functions of stellar mass $M_*$ of the central galaxy through means described in the previous section. As can be seen on the left panel, at a given mass ratio $\mu=m/M$, the infall time-scale falls very weakly across a wide range of stellar masses. As the right panel shows, the time-scale varies strongly for $\mu<0.25$, with values between $\mu=0.25$ and $\mu=1$ changing only weakly. 

\section{Stellar mass growth rate}
\label{sec:section4}

Given an infall time-scale formula and the satellite distribution around a given galaxy, the calculation of the {\it expected} stellar mass growth rate (hereafter: SMGR) is made possible. We begin by associating an average growth rate of a galaxy of stellar mass $M_*$ due to the infall of a single subhalo of mass $m$ hosting a galaxy of stellar mass $m_*$:
\begin{equation}
    \frac{\Delta M_*}{\Delta t}= F(m_*,\tau).
	\label{eq:GrowthToOne}
\end{equation}
We leave the function $F$ unspecified for the moment.

The SMGR of a galaxy of stellar mass $M_*$ is found by summing over all individual growth rates (given by equation \ref{eq:GrowthToOne}), multiplied by the expected mass and spatial distribution:
\begin{equation}
    \frac{\diff M_*}{\diff t}=\int \int \frac{\diff N}{\diff m \diff V} F(m_*,\tau)\diff m \diff V,
	\label{eq:SMGRGeneral}
\end{equation}
where $\diff N/\diff m \diff V$ is the number distribution of subhaloes with respect to their masses and positions around the central galaxy. The SMGR, given in this form, is general and can be applied in different models with different infall time-scales $\tau$.

Since the probability of finding a subhalo at a given distance $r$ from the centre of the parent halo is independent of subhalo mass $m$ \cite{Han2015}, the distribution $\diff N/\diff m \diff V$ can be seperated onto its mass and spatial components,
\begin{equation}
    \frac{\diff N}{\diff m \diff V}=P(r)\frac{\diff N}{\diff m},
	\label{eq:Distribution}
\end{equation}
where $\diff N/\diff m$ is the halo mass function (the number density of subhalos in a mass interval $m\pm \diff m/2$), an extensively studied quantity, while $P(r)$ is the probability density of finding a subhalo at a point $(r,\theta,\phi)$. 

In general, tidal stripping alters the subhalo spatial distribution, but this does not change the probability distribution of finding {\it satellites} at a given position \cite{Guo2013, Sales2007, Han2015}. They follow their initial distribution, which is the same as the dark matter distribution of the halo itself. The probability density P(r) is then given by the NFW profile,
\begin{equation}
   P(r)=\frac{P_0}{x(1+x)^2},
	\label{eq:SpatialProbDensity}
\end{equation}
with $x=r/r_s$. We determine the normalisation $P_0$ through the requirement that the satellite must be found somewhere between the center of its parent halo and the virial radius.  In other words, the volume integral of $P(r)$ within the host halo must equal unity. From this condition, $P_0$ follows as
\begin{equation}
P_0=\frac{1}{r_s^3[\ln(1+c)-c/(1+c)]}=\frac{4\pi \rho_0}{M},
	\label{eq:SpatialProbNorm}
\end{equation}
where the second equality follows from the formula for enclosed mass at virial radius (equation \ref{eq:EnclosedMass}).

The number density $dN/dm$ of subhaloes with respect to mass (hereafter: subhalo mass function; SHMF) is generally modelled by the form
\begin{equation}
    \frac{\diff N}{\diff m}=\frac{N_1}{M}\mu^{-\alpha}\exp\Big(-C\mu^{\beta}\Big),
	\label{eq:SMF}
\end{equation}
with $\mu=m/M$ the subhalo mass in units of parent halo mass, and $N_1$ the normalisation. The exponent $\alpha$ varies from simulation to simulation, but is generally constrained to the range [1.8, 2] \cite{Bosch2005, Giocoli2010, Han2015}. Some models do not incorporate an exponential high-mass cutoff \cite{Han2015}, while among those that do, the exponent $\beta$ is not unique. The authors of \cite{Rodriguez-Puebla2013} use $\beta=1.3$ and $\alpha=1.96$, while those of \cite{Giocoli2010} set $\beta=3$, $\alpha=1.8$, with the factor $C=12.27$. We follow the latter model.

\begin{figure}
\centering
	\includegraphics[width=0.6\columnwidth]{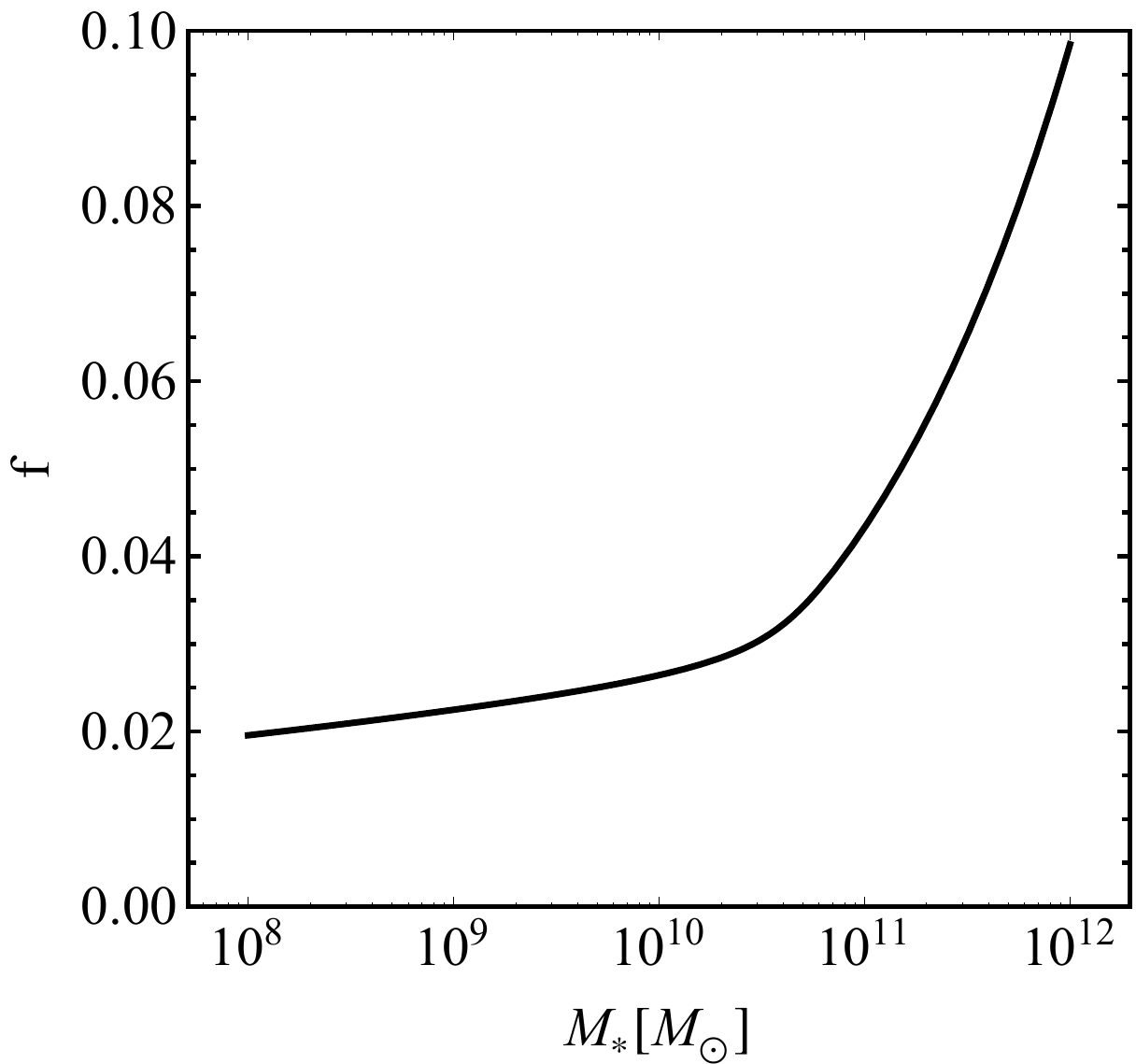}
    \caption{Total mass fraction of parent halo contained in subhaloes as a function of central galaxy stellar mass.}
    \label{fig:fig3}
\end{figure}

The total mass contained in subhaloes is generally well determined \cite{Bosch2005}. This can be used to find the normalisation $N_1$,
\begin{equation}
    N_1\int_{m_{\textnormal{min}}}^{m(M_*)}m\frac{\diff N}{\diff m} \diff m=fM,
	\label{eq:Normalisation}
\end{equation}
where we take the minimal mass to be integrated $m_{\textnormal{min}}=10^{-4}M$ (the mass resolution in simulations done in \cite{Bosch2005}), while the upper limit $m(M_*)$ is the mass of a subhalo that hosts a satellite equally massive as the central galaxy. This would be simply equal to the primary halo mass $M$ if not for tidal stripping. As found in \cite{Bosch2005}, the total mass fraction $f$ of subhaloes contained in the parent halo (above the mass resolution limit) is a function of parent halo mass.  More massive haloes have more of their mass contained in subhaloes. We model the findings of \cite{Bosch2005} by the following equation:
\begin{equation}
    f(M,z)=0.05\bigg( \frac{M}{\textnormal{M}_{\sun}}\bigg)^{0.14}\sqrt{1+z}.
	\label{eq:MassFraction}
\end{equation}
The redshift dependence was taken from \cite{Giocoli2010}. The mass fraction contained in subhaloes is shown on Fig. \ref{fig:fig3} as a function of central galaxy stellar mass through the relation $M(M_*)$ (at redshift $z=0$).

The normalisation $N_1$ now follows as
\begin{equation}
   N_1(M_*,z)=f(M,z)\bigg/ \int_{10^{-4}}^{\mu(M_*)}\mu^{-0.8}e^{-12.27\mu^3}\diff \mu,
	\label{eq:NormalisationFinal}
\end{equation}
where the upper limit $\mu=m(M_*)/M(M*)$ is the mass ratio of a subhalo that hosts a satellite equally massive as the central galaxy and the parent halo of the galaxy. The integral itself was obtained from equation (\ref{eq:Normalisation}) by a simple substitution, $m=\mu M$.

\subsection{Model I: Instantaneous merging}
\label{sec:subsection4.1}

With both the spatial and mass distributions of subhaloes determined, we now turn to the function $F$ that determines the individual subhalo contribution to the SMGR (equation \ref{eq:GrowthToOne}). In our first model, we assume that mergers of stellar components are instantaneous. This is in accord with \cite{Chang2013}, where it was found that the stellar mass $m_*$ of a satellite is stripped rather violently after the host subhalo loses 90\% of its mass. Mathematically, this means that the function $F_\textnormal{I}$ must be proportional to the Dirac delta in time:
\begin{equation}
  \frac{\Delta M_*}{\Delta t}= F_\textnormal{I}(m_*,\tau)=\langle m_*\rangle \delta(t-\tau),
	\label{eq:F_I}
\end{equation}
It should be noted that this individual growth rate contributes to the SMGR at time $t$, not to the current ($t=0$) SMGR.

\begin{figure}
\centering
	\includegraphics[width=0.6\columnwidth]{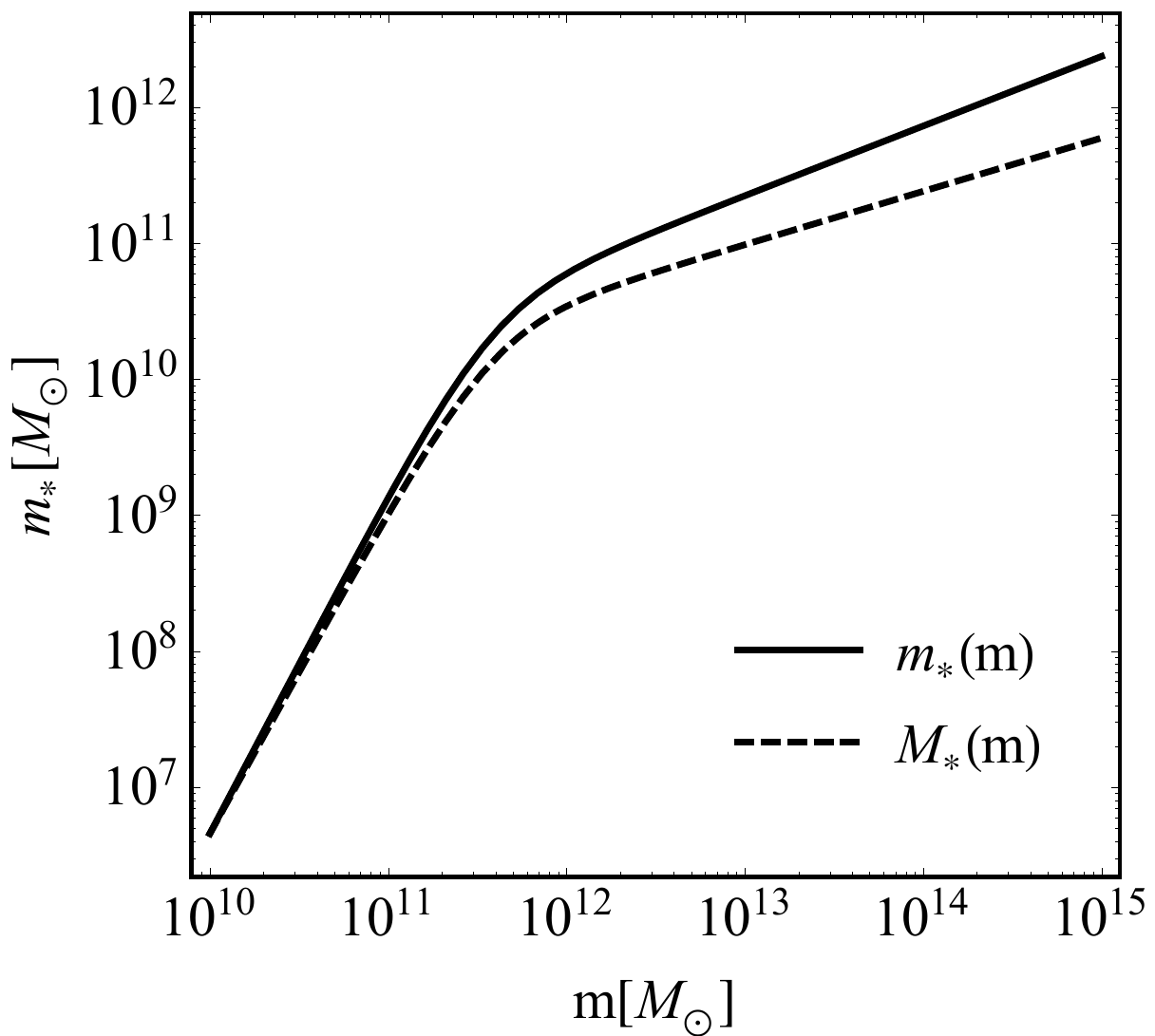}
    \caption{Satellite galaxy stellar mass as a function of its host subhalo mass. The dashed line shows (for comparison) the stellar mass the satellite would have if its stellar mass was given by the central galaxy SHMR.}
    \label{fig:fig4}
\end{figure}

Determining the stellar mass $m_*(m)$ of a satellite will also determine (through its inverse) the upper limit of the normalisation $N_1$ (equation \ref{eq:NormalisationFinal}). The satellite stellar-to-subhalo mass relation $m_*(m)$ is generally similar to the one for central galaxies, $M_*(M)$ (because satellite galaxies were once field galaxies) \cite{Moster2010}. However, the satellite SHMR is larger at a given (sub)halo mass, as caused by tidal stripping. This discrepancy increases with (sub)halo mass (Fig. 8 in \cite{Moster2010}, top right panel). Specifically, the two relations overlap completely at (sub)halo mass of $10^{10}$ M$_{\sun}$, and differ by a factor of around $4$ at $10^{15}$ M$_{\sun}$. We model this by the following relation:
\begin{equation}
   m_*(m)=\bigg(\frac{m}{10^{10}\textnormal{M}_{\sun}} \bigg)^{0.11}M_*(m),
	\label{eq:SatSHMR}
\end{equation}
where $M_*(m)$ is the central galaxy SHMR, as given by equation (\ref{eq:SHMR}), while the additional factor accounts for the difference between the two SHMRs. The redshift dependences are inherited from the central galaxy SHMR. Fig. \ref{fig:fig4} shows the two SHMRs.

The expected stellar mass in equation (\ref{eq:F_I}) contained in a subhalo of mass $m$ differs somewhat from the SHMR; $\langle m_*(m)\rangle\neq m_*(m)$. This is due to the fact that the SHMR assumes that the subhalo hosts a galaxy, while this may not necessarily be the case. The integration in the SMGR is done over all subhaloes, which means that the expected stellar mass $\langle m_*(m)\rangle$ must factor in those subhaloes that host no galaxies. We account for this by assuming the expected stellar mass to be given simply by 
\begin{equation}
   	\langle m_*(m)\rangle=p \hspace{0.3mm}m_*(m),
	\label{eq:Expected SHMR}
\end{equation}
where $p$ is the mean probability that a subhalo hosts any satellite at all. This parameter is related to the `missing satellites' problem. Namely, simulations overpredict the number of satellites as compared to observational studies. We fixate $p=0.1$ hereafter, which is the value that `fixes' this problem \cite{Nickerson2011}. 

With the expected stellar mass determined, we now substitute the individual growth rate (equation \ref{eq:F_I}) into our general SMGR formula (equation \ref{eq:SMGRGeneral}). Choosing the spatial integral as the inner one, we have
\begin{equation}
   	\frac{\diff M_*}{\diff t}(t)=\int \diff m \langle m_* \rangle \frac{\diff N}{\diff m}\int \diff r P(r)r^2\delta(t-\gamma r^2),
	\label{eq:SMGRI1}
\end{equation} 
where we have already substituted the $r$-depedence of $\tau$; $\tau=\gamma r^2$. Here, $\gamma$ reads (from equation \ref{eq:FinalTimescale}) as
\begin{equation}
   	\gamma=\frac{16}{5}\sqrt{\frac{\rho_0}{G}}\frac{r_s}{m\ln(1+M/m)}.
	\label{eq:Gamma}
\end{equation}
The delta function in the spatial integrand is quadratic in $r-$it can be seperated onto two delta functions. Specifically, we have
\begin{equation}
   	\delta(t-\gamma r^2)=\frac{1}{2\sqrt{\gamma t}}\bigg[ \delta\bigg(r-\sqrt{\frac{t}{\gamma}}\bigg)+\delta\bigg(r+\sqrt{\frac{t}{\gamma}}\bigg) \bigg],
	\label{eq:Delta}
\end{equation}
which follows from basic properties of the delta function. The second of the two delta functions is irrelevant ($r$ would need to be negative). The spatial integrand is then simply equal to $P(r)r^2$ evaluated at $r=\sqrt{t/\gamma}$. This leads to
\begin{equation}
   	\frac{\diff M_*}{\diff t}(t)=\frac{1}{2}P_0\int_0^{m(\alpha M_*)} \diff m \langle m_* \rangle \frac{\diff N}{\diff m} \frac{r_s}{\gamma(1+\sqrt{t}/\sqrt{\gamma} r_s)^2}.
	\label{eq:SMGRI2}
\end{equation}
The upper bound of the integral is parametrized in terms of the stellar mass merger ratio $\alpha$. In other words, the most massive subhalo to contribute to the SMGR is that whose stellar mass is a fraction $\alpha$ of the central galaxy. In the above equation, we can see that taking the limit $t\rightarrow0$ (which is what we're interested in) would give a finite result. Since the limit of a sum is equal to a sum of limits, we can apply the limit to the integrand directly ($\gamma$ depends on $m$ and cannot be taken out of the integral). Taking this limit, as well as substituting in $\gamma$, writing out $P_0$ and the SHMF $\diff N/\diff m$, we finally have the following formula for the SMGR in the instantaneous merger model:
\begin{equation}
   	\frac{\diff M_{*,I}}{\diff t}=\frac{5\pi}{8}pN_1\sqrt{G\rho_0}MI(M_*,\alpha),
   		\label{eq:SMGRIFinal}
\end{equation}
where the function $I$ is given by
\begin{equation}
   	I(M_*,\alpha)=\int_{10^{-4}}^{\mu_{\alpha}} \mu^{0.2} e^{-12.27\mu^3} \ln \Big(1+\frac{1}{\mu}\Big)\frac{m_*(\mu M)}{\mu M}.
   		\label{eq:MassIntegral}
\end{equation}
This was obtained from the mass integral through a simple substitution $m=\mu M$. Thus, $\mu$ represents the (sub)halo merger ratio. The upper bound of the integral $\mu_\alpha$, which represents the ratio of halo masses whose central galaxies have a stellar mass ratio equal to $\alpha$, is a complicated function of stellar mass $M_*$, and is given by
\begin{equation}
   	\mu_\alpha=\frac{m(\alpha M_*)}{M(M_*)}.
	\label{eq:MuAlpha}
\end{equation}

\subsection{Model II: Continuous merging}
\label{sec:subsection4.2}

In this model we assume that satellites are stripped continuously. This essentially corresponds to galaxies growing through accretion from their satellites, although we still refer to this process as merging.

We choose the form of the individual growth rate (function $F$) in the simplest possible manner. Namely, we assume that the accretion rate is constant over the entire infall time-scale $\tau$. This corresponds to
\begin{equation}
  \frac{\Delta M_*}{\Delta t}= F_\textnormal{II}(m_*,\tau)=\frac{\langle m_*\rangle}{\tau(m,r)}.
	\label{eq:F_II}
\end{equation}
At first, this seems somewhat unrealistic. For example, looking at Fig. \ref{fig:fig2} one can see that infall time-scales can sometimes be up to several Gyr long, which means that they do not contribute to the current SMGR. While this is true, there are two mechanisms which suppress the contribution of these problematic satellites to the current SMGR. 

Firstly, the infall time-scales shown on Fig. \ref{fig:fig2} correspond to subhaloes located at the edge of the host halo. However, satellite galaxies are spatially distributed according to the NFW profile, which has a maximum (once the growth due to spherical shells as $r^2$ is included) at $r/r_s=1$. Host haloes have concentrations (sizes in units of $r_s$) that fall between $c=5$ and $c=15$ for galaxies in the $10^{8}$ M$_{\sun} - 10^{12}$ M$_{\sun}$ interval. Since the infall time-scale grows as $r^2$, this means that most satellite galaxies have infall time-scales well below those shown on Fig. \ref{fig:fig2}$-$the number of problematic satellites is relatively negligible. Secondly, since these satellites have by definition larger infall time-scales, their contribution to the SMGR through equation (\ref{eq:F_II}), which goes as $1/r^2$, is automatically smaller.

\begin{figure}
\centering
	\includegraphics[width=0.6\columnwidth]{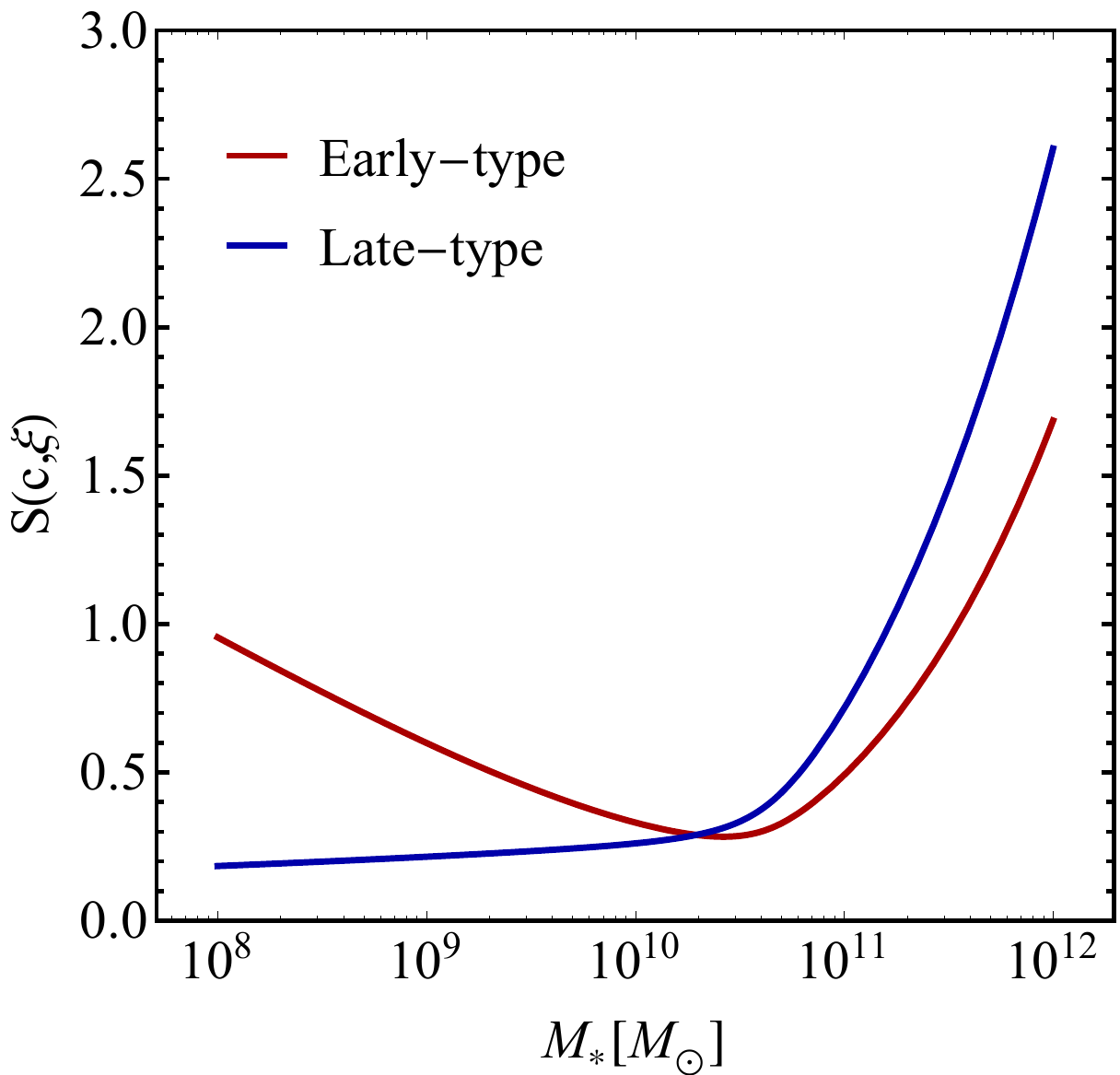}
    \caption{Spatial factor in the stellar mass growth rate of galaxies, given by equation (\ref{eq:SpatialFunction}), as a function of central galaxy stellar mass $M_*$ for both early and late-type galaxies.}
    \label{fig:fig5}
\end{figure}

Substituting this form of $F$ in our general SMGR formula, we arrive at an immediately seperable integral:
\begin{equation}
    \frac{\diff M_*}{\diff t}=\frac{5}{16}\sqrt{\frac{G}{\rho_0}}\frac{1}{r_s^3} \int \frac{P(r)}{(r/r_s)^2}\diff V \int\frac{\diff N}{\diff m}m\langle m_*(m)\rangle \ln \Lambda \diff m ,
	\label{eq:SMGRII1}
\end{equation}
with $\Lambda=1+M/m$. The lower boundary of the spatial integral is chosen to be the perifery of the parent galaxy (determined by the GH ratio $R/r_s$, as discussed in Section \ref{sec:subsection2.3}). With this choice, we are effectively only considering those satellites that do not contribute to the observed luminosity of the central galaxy. The upper boundary of the integral is taken to be the radius of the host halo. The spatial integral can then be evaluated analytically to yield
\begin{equation}
  S(c,\xi)=\frac{\ln[c(1+\xi)/\xi(1+c)]+1/(1+c)-1/(1+\xi)}{\ln(1+c)-c/(1+c)},
	\label{eq:SpatialFunction}
\end{equation}
where $\xi=R/r_s$ is the GH ratio and $c$ the parent halo concentration. This function is shown on Fig. \ref{fig:fig5} for both early and late-type galaxies. Early-type galaxies are expected to grow faster in the lower mass regime, while late-types are expected to grow faster in the high-mass regime. This does not predict that high-mass disk galaxies should be common. In fact, it predicts the opposite. The more prone these galaxies are to merging, the more prone their disks are to being disrupted.

With the spatial integral determined (the mass integral is the same as in model I), the SMGR is  given by
\begin{equation}
   	\frac{\diff M_{*,II}}{\diff t}=2S_0(c,\xi)\frac{\diff M_{*,I}}{\diff t}=\frac{5\pi}{4}pN_1M\sqrt{G\rho_0}I(M,\alpha)S_0(c,\xi),
	\label{eq:SMGRIIFinal}
\end{equation}
where $S_0(c,\xi)$ is the numerator of the spatial integral (equation \ref{eq:SpatialFunction}):
\begin{equation}
  S_0(c,\xi)=\ln[c(1+\xi)/\xi(1+c)]+1/(1+c)-1/(1+\xi).
	\label{eq:SpatialFunctionNum}
\end{equation}

\section{Results and analysis}
\label{sec:section5}

The parameters, $\rho_0$, $c$ and $N_1$ in our two SMGR equations (\ref{eq:SMGRIFinal} and \ref{eq:SMGRIIFinal}) are all dependent on halo mass $M$ (and thus the central galaxy mass $M_*$) as well as redshift $z$, with $\xi$ being dependent on $M_*$ and $z$ through its definition $\xi=R/r_s$. The probability $p$ of subhaloes hosting satellites is taken to be 0.1, as discussed in Section \ref{sec:subsection4.1}. At a given stellar mass $M_*$ and redshift $z$, the cumulative mass growth rate due to all mergers of ratio smaller than $\alpha$ is then determined by the derived formulas.

\subsection{Comparison with simulations at $z=0$}
\label{sec:subsection5.1}

In order to test the validity of our derived SMGRs, as well as make a comparison between the two, we consider the total SMGRs, characterised by the maximal merger stellar mass ratio $\alpha=1$. This corresponds to choosing the upper limit of the mass integral to be

\begin{equation}
   	 \mu_\textnormal{max}=\mu(\alpha=1)=\frac{m(M_*)}{M(M_*)}.
	\label{eq:UpperMassLimit}
\end{equation}

\begin{figure}
\centering
	\includegraphics[width=0.6\columnwidth]{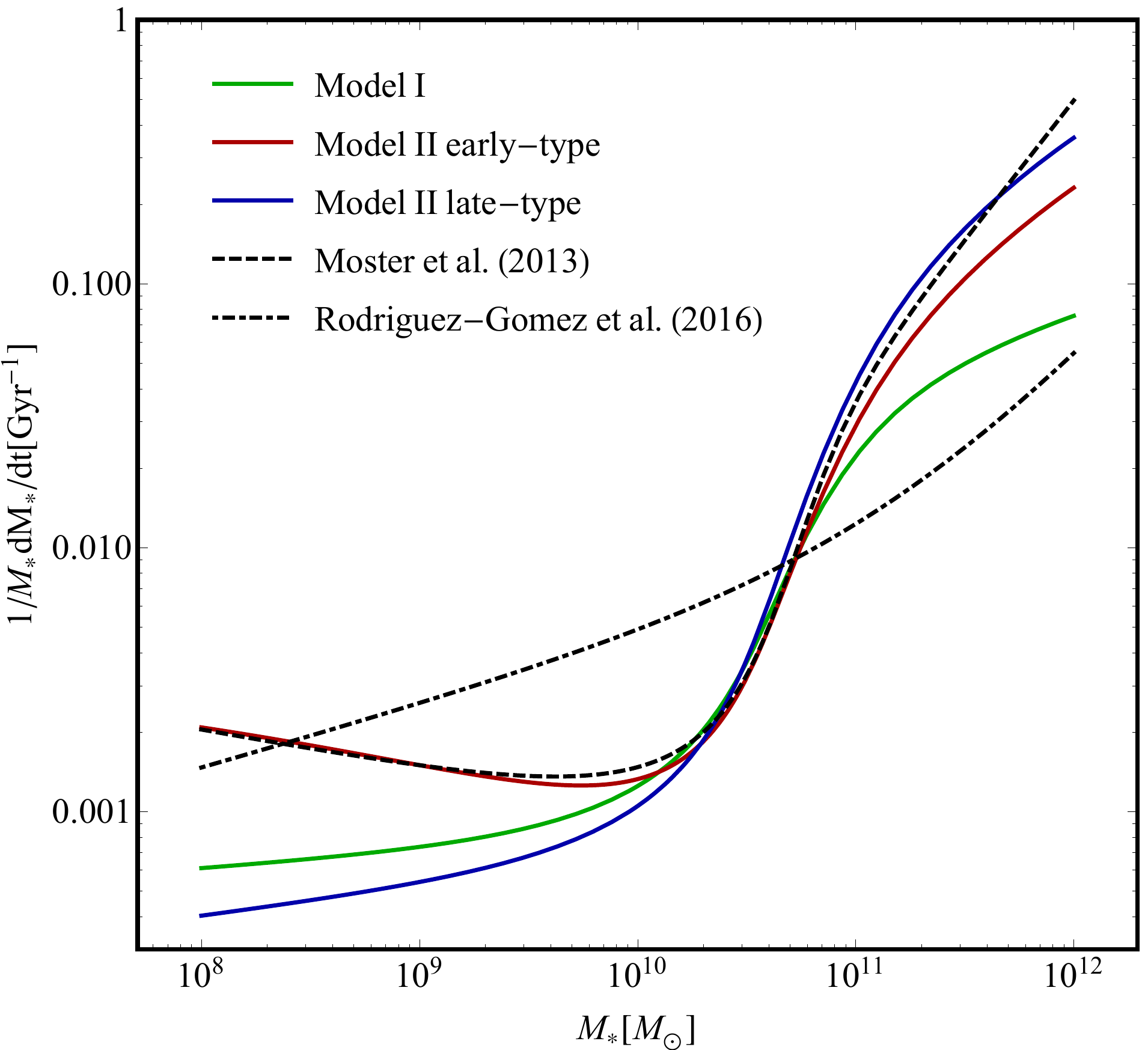}
    \caption{Specific stellar mass growth rate as a function of central galaxy stellar mass. Colours represent our two different models (and two galaxy types in model II), according to the legend. The dashed line represents the results from the Millennium-II simulation as analyzed in \protect\cite{Moster2013}. The dot-dashed line is the result from the Illustris simulation, discussed in \protect\cite{Rodriguez-Gomez2016}. }
    \label{fig:fig6}
\end{figure}

We compare our SMGRs with fits found in two recent cosmological simulations. A fitting formula based on the Millennium-II simulation can be found in \cite{Moster2013}, while the authors of \cite{Rodriguez-Gomez2016} offer a formula based on the Illustris simulation. Fig. \ref{fig:fig6} shows this comparison, with the {\it specific} SMGR (SMGR divided by $M_*-$this quantity measures the growth rate in units of current stellar mass) plotted against central galaxy stellar mass $M_*$. 

Our relations follow the results of the Illustris simulation (dot-dashed line) only in a rough sense. However, the same simulation also predicts that major mergers contribute about 60\% to total merger growth across all stellar masses \cite{Rodriguez-Gomez2016}, which does not agree with observation. We discuss this further in Section \ref{sec:subsection5.3}.

Our model I (instantaneous merging) agrees with the Millennium-II simulation only in the $10^{10}$ M$_{\sun} - 10^{11}$ M$_{\sun}$ interval. For smaller and larger galaxies, the model underpredicts the growth due to mergers. Thus, accretion from satellites seems to be more important in these regimes. Model II agrees with the results of the simulation, although the low-mass end of the SMGR is only reproduced by our early-type model II relation. There is a small disagreement between our results and the simulation at the very high mass end, although this is expected since our model assumes that the host galaxy is isolated  rather than being the central galaxy of a cluster. 

\subsection{Redshift evolution}
\label{sec:subsection5.2}

The redshift dependences of our SMGR formulas offer an additional method for testing their validity. Here we compare the redshift evolution of our SMGR in model II (continuous merging) with results from the Millennium-II simulation \cite{Moster2013}, since model II represents the corresponding SMGR at $z=0$ correctly. We did not assume any redshift evolution of effective galactic radii (and thus galactic sizes) in accord with \cite{Barden2005}, where it was found that the size evolution is negligble for small $z-$we restrict ourselves to this regime. 

\begin{figure*}
	\includegraphics[width=\textwidth]{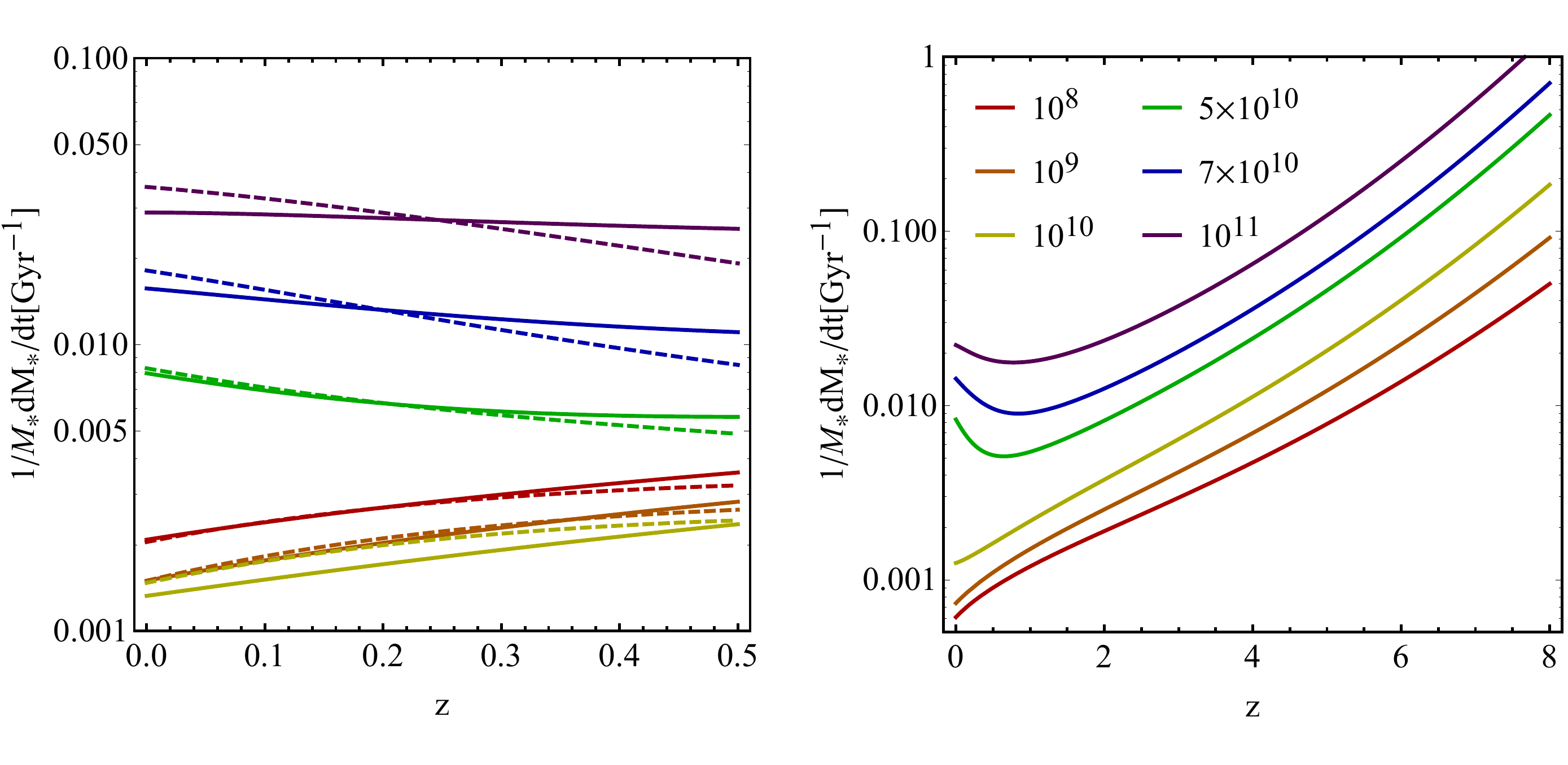}
    \caption{Redshift evolution of the early-type specific stellar mass growth rate. {\it Left}: Evolution of the model II SMGR (full lines) for small redshift, in comparison with the results of the Millennium-II simulation \protect\cite{Moster2013} (dashed lines). {\it Right}:  High-redshift evolution of our model I SMGR. Different central galaxy stellar masses are shown by varying colours as given by the legend (the units being solar masses). The colours follow the spectrum sequence such that a redder colour represents a lower mass galaxy.}
    \label{fig:fig7}
\end{figure*}

The left panel of Fig. \ref{fig:fig7} shows the redshift evolution (for $z<0.5$) of the model II early-type specific SMGR for several stellar masses. Low and intermediary mass galaxies ($10^8$ M$_{\sun} - 5\times10^{10}$ M$_{\sun}$) exhibit a redshift evolution in rough agreement with the Millennium-II siulation (dashed lines) \cite{Moster2013}, while high mass galaxies have a larger SMGR with increasing redshift than predicted with our formula. This discrepancy is not surprising due to the relatively strong size evolution of high-mass ellipticals since $z=1$ \cite{Huertas-Company2013}, while we assumed size to be constant even for these galaxies.

At higher redshifts, we expect that mergers become relatively instantaneous and violent, with infall time-scales being much shorter. Due to this, our model I should give the correct SMGR in this regime. This is shown on the right panel of Fig. \ref{fig:fig7}. The lower redshift evolution should be ignored. After a certain redshift, the evolution becomes mass-independent (with the exception of the normalisation). Specifically, the SMGR grows as $(1+z)^4$ at these times. This kind of behaviour (growing SMGR rather than waning) is expected in the hierarchial merging picture of galaxy formation.

\subsection{Merger type contributions}
\label{sec:subsection5.3}

First we consider the two usual merger types; minor mergers ($\alpha<0.25$) and major mergers ($0.25<\alpha<1$). We define the relative contribution $\Phi$ of each of the two types as the fraction of the total SMGR caused by merging events of the corresponding type. This is given by
\begin{equation}
   	\Phi_{\textnormal{min}}=\frac{SMGR(0.25)}{SMGR(1)},
	\label{eq:MergerTypeImpact}
\end{equation}
for minor mergers, where $SMGR(\alpha)$ is the cumulative SMGR due to all merger ratios smaller than $\alpha$, given by either of our two SMGR equations (\ref{eq:SMGRIFinal} or \ref{eq:SMGRIIFinal})$-$since the two models do not differ in the mass integral, it is irrelevant which equation we use. We reiterate that given a specific $\alpha$, the upper boundary $\mu_\alpha$ in the mass integral in the SMGR is generally different than $\alpha$ and is a function of $M_*$. The relative contribution for major mergers can be found simply as 
\begin{equation}
   	\Phi_{\textnormal{maj}}=1-\Phi_{\textnormal{min}}.
	\label{eq:MergerTypeImpact}
\end{equation}
By its very definition, the relative contribution does not depend on any of the following parameters: $\rho_0$, $r_s$, $c$, $f$, nor the spatial function $S_0(c,\xi)-$they all cancel out. Thus $\Phi$ depends only on stellar mass $M_*$ and redshift $z$ through the stellar-to-halo mass relation and its inverse, in both its variants (for central and satellite galaxies). These are valid for $z<4$ \cite{Moster2013}. In effect, we can study the relative merger contributions for large $z$ (up to $z=4$) even though we cannot study the SMGR itself.

\begin{figure*}
	\includegraphics[width=\textwidth]{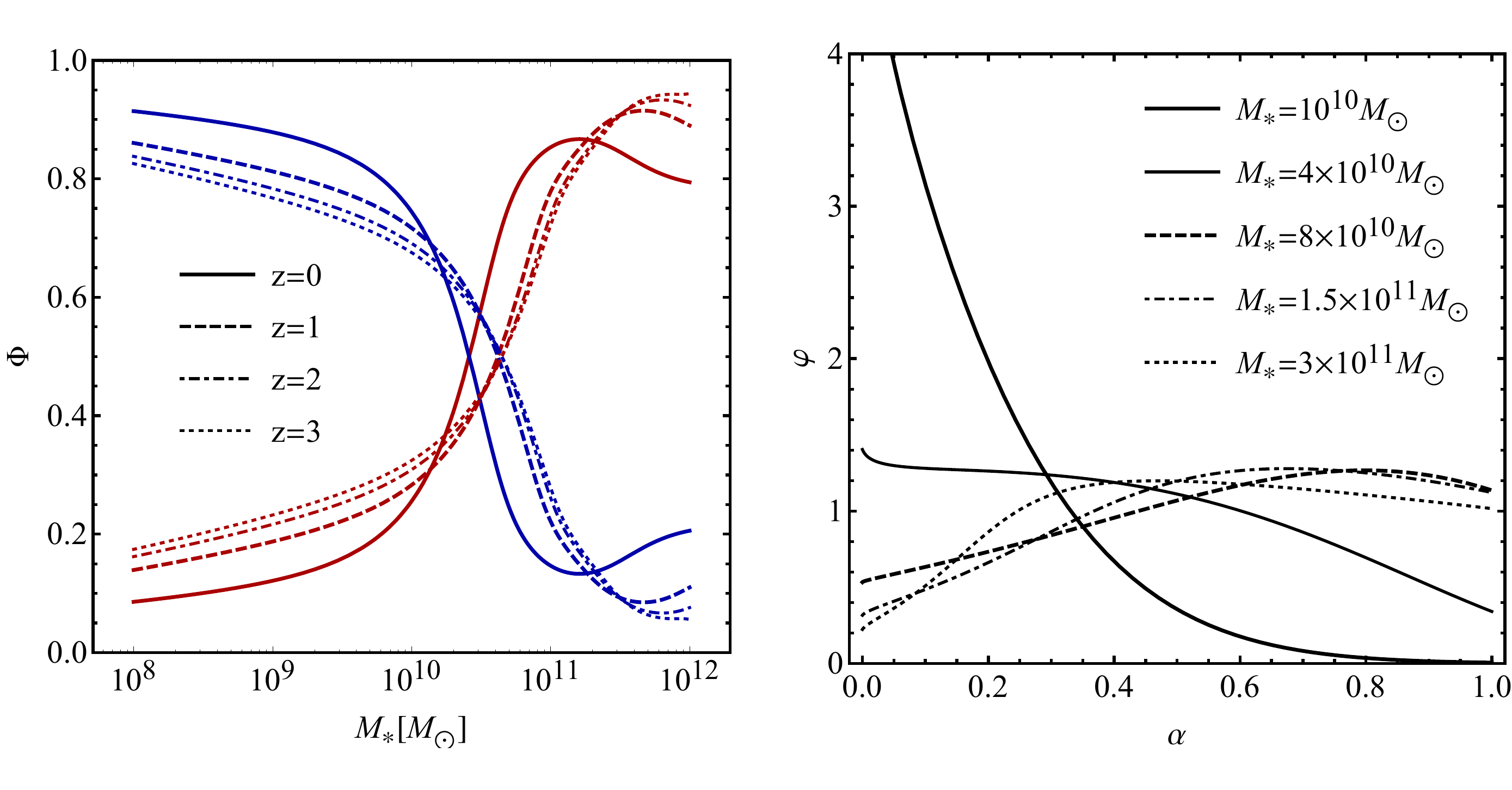}
    \caption{Relative contributions of merger types to the total stellar mass growth rate. {\it Left}: Relative contributions of minor and major mergers as functions of central galaxy stellar mass at four different redshifts, as given by the legend. The blue and red lines represent minor and major merger contributions respectively. {\it Right}: Merger ratio contribution density as a function of merger ratio $\alpha$ for different central galaxy stellar masses (according to the legend) at redshift $z=0$. }
    \label{fig:fig8}
\end{figure*}

The left panel of Fig. \ref{fig:fig8} shows the relative contribution of minor and major mergers at four different redshifts from $z=0$ to $z=3$. Minor mergers dominate in the lower mass range ($10^{8}$ M$_{\sun} - 10^{10}$ M$_{\sun}$), where they contribute 80\% to 90\% of the total SMGR, with this proportion increasing with decreasing redshift. In the range $10^{10}$ M$_{\sun} - 10^{11}$ M$_{\sun}$ the impacts reverse, with major mergers becoming the dominant process. The existence of a local maximum exhibited in the major merger contribution (and the resurgence of minor mergers as an important process at high masses) can be explained through the relative rarity of high-mass galaxies. An additional factor in this complex interplay is the existence of a maximum in the stellar-to-halo mass ratio at around the Milky Way stellar mass. The existence of this maximum and its position (on the stellar mass axis) are in agreement with previous studies on the relative importance of merger types \cite{Hopkins2010b, Cattaneo2011}.

Our results are in good agreement with observation. The authors of \cite{Bundy2009} find that the transition (in dominance) from minor mergers to major mergers occurs between $10^{10}$ M$_{\sun}$ and $10^{11}$ M$_{\sun}$, which our analysis reproduces nicely. The point where the contribution of minor and major mergers becomes equal (in our results) is found at a relatively high mass ($\approx 4\times 10^{10}$ M$_{\sun}$) for redshift $z>1$, which is to say that minor mergers are expected to have been the dominant merger type in galaxy formation and early evolution$-$this is in agreement with previous findings \cite{GuoWhite2008, Parry2009, Oogi2013, Kaviraj2014}. The non-vanishing contribution of major mergers at low masses (10\%$ - $20\%) agrees with observations, which find that dwarf galaxies do experience such mergers \cite{Amorisco2014, Deason2014, Koch2015}.

While minor and major mergers are the usual categories of interest, our model allows a more general study. We define the merger ratio contribution {\it density} as
\begin{equation}
   	\varphi(\alpha)=\frac{1}{SMGR(1)}\frac{\diff SMGR(\alpha)}{\diff \alpha},
	\label{eq:MergerRatioImpactDensity}
\end{equation}
where $SMGR(\alpha)$ is again the cumulative SMGR due to all merger ratios smaller than $\alpha$. Defined in this way, $\varphi$ measures the contribution of mergers of ratio $\alpha$ to the total SMGR. The right panel of Fig. \ref{fig:fig8} shows this quantity (at redshift $z=0$) as a function of $\alpha$ for stellar masses in the range $10^{10}$ M$_{\sun} - 3\times10^{11}$ M$_{\sun}$, where the relative contributions of minor and major mergers reverse. At $M_*=10^{10}$ M$_{\sun}$ (and for stellar masses smaller than this), the merger contribution density is a decreasing function of $\alpha$, with larger merger ratios contributing ever less to the SMGR. At $M_*=4\times10^{10}$ M$_{\sun}$, the density is approximately constant for $\alpha$ between $0$ and $0.5$.  At $M_*=8\times10^{10}$ M$_{\sun}$, a maximum appears near $\alpha=0.8$. At higher masses, this maximum shifts to the left, showing the recurrence of minor mergers as an important process. Very minor mergers ($\alpha<0.1$), however, contribute less and less with increasing stellar mass. This diminishing contribution is in agreement with Millennium data \cite{Oogi2013}.

\subsection{Hierarchal formation of galaxies}
\label{sec:subsection5.4}

Having shown in the previous subsections that our model is consistent with simulations and observations in the near and late universe, here we aim to show that it is also consistent with the picture of hierarchal formation of galaxies. We expect that mergers in the early universe are relatively violent; the appropriate SMGR that can describe the growth of a central protogalaxy is the one laid out in Section \ref{sec:subsection4.2} (model I SMGR). However, this needs to be modified slightly. The final form of the model I SMGR is given by
\begin{equation}
   	\frac{\diff M_{*}}{\diff t}=\frac{5\pi}{8}pN_1\sqrt{G\rho_0}MI(M_*,\alpha)
   		\label{eq:SMGRIFinal2}
\end{equation}
For the early universe, it is reasonable to assume that all subhaloes host galaxies$-$this sets $p$ equal to unity. We now look at the factor $N_1\hspace{0.3mm}I(M_*, \alpha)$:
\begin{equation}
   	N_1\hspace{0.3mm}I(M_*,\alpha)=f \int_{10^{-4}}^{\mu_{\alpha}} \mu^{0.2} e^{-12.27\mu^3} \ln \Big(1+\frac{1}{\mu}\Big)\frac{m_*(\mu M)}{\mu M}\bigg/ \int_{10^{-4}}^{\mu_{\alpha}}\mu^{-0.8}e^{-12.27\mu^3}\diff \mu.
   		\label{eq:MassIntegral2}
\end{equation}
We assume the maximal mass fraction $\mu_\alpha$ is also equal to unity for the early universe (since galaxy fragments are all assumed to be similar). Furthermore, we replace the stellar mass $m_*$ in the above integral with the baryonic mass $m_\textnormal{b}$, which has a much simpler dependence on halo mass $M$:
\begin{equation}
   	m_*(M)=f_\textnormal{b}(1-f_\textnormal{CW})M,
   		\label{eq:SHMRearlyuniverse}
\end{equation}
where $f_\textnormal{b}=0.186$ is the cosmic baryon fraction \cite{Ade2015}, while $f_\textnormal{CW}=0.5$ is the baryon fraction contained in the cosmic web \cite{Eckert2015}. Taking all of these assumptions into account, equation (\ref{eq:MassIntegral2}) simplifies to
\begin{equation}
   	N_1\hspace{0.3mm}I(M_*,\alpha)=f f_\textnormal{b} (1-f_\textnormal{CW}) \int_{10^{-4}}^{1} \mu^{0.2} e^{-12.27\mu^3} \ln \Big(1+\frac{1}{\mu}\Big)\bigg/ \int_{10^{-4}}^{1}\mu^{-0.8}e^{-12.27\mu^3}\diff \mu.
   		\label{eq:MassIntegral22}
\end{equation}
Both of the above integrals can now be evaluated numerically. Their ratio is within 0.7\% of $\pi/20$, so we choose this value for sake of simplicity. 

With these integrals evaluated, our early-universe galaxy growth rate is equal to
\begin{equation}
   	\frac{\diff M_{\textnormal{b}}}{\diff t}=\frac{\pi^2}{32}\sqrt{G\rho_0}M_{\textnormal{b,\hspace{0.2mm}sat}},
   		\label{eq:EarlyUniverseGrowthRate}
\end{equation}
where $M_{\textnormal{b,sat}}=f f_\textnormal{b} (1-f_\textnormal{CW})M$ is the total baryonic mass contained in satellites of the central galaxy. On the left hand side, the stellar mass $M_*$ has been replaced with the baryonic mass $M_\textnormal{b}-$this step is necessary since we have also replaced stellar masses with baryonic masses while integrating over all subhaloes in equation (\ref{eq:MassIntegral2}). In effect, our new formula measures the baryon accretion rate onto the central galaxy due to mergers.

Equation (\ref{eq:EarlyUniverseGrowthRate}) is essentially a differential equation for the unkown function $M_\textnormal{b}(z)$. We solve this equation numerically. The inital condition is $M_\textnormal{b}(z_i)=10^6$ M$_\sun$. Here, $10^6$ M$_\sun$ is the mass of overdensities collapsing in on themselves after recombination (initial protogalaxy mass) \cite{Carroll2007}, while $z_i$ represents the redshift at which this overdensity collapses and forms a protogalaxy. We treat the initial mass as constant and equal at all initial redshifts, while initial redshift is the variable that distinguishes modern galaxies of differing masses.

\begin{figure}
\centering
	\includegraphics[width=0.6\columnwidth]{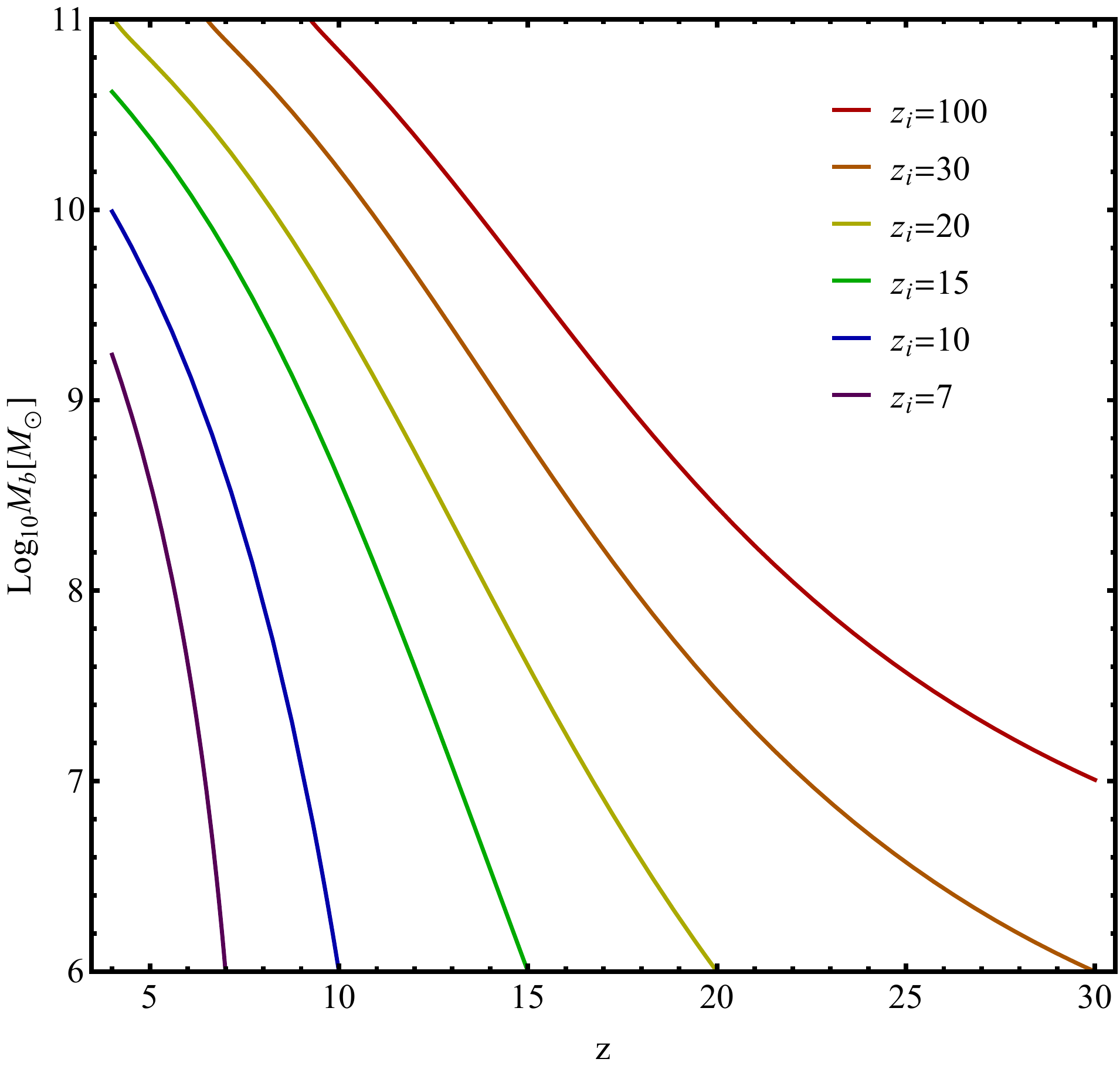}
    \caption{Growth history of galaxy baryonic content in the early universe. Varying colours represent galaxies that formed and started growing earlier, according to the legend.}
    \label{fig:fig9}
    \end{figure}

Fig. \ref{fig:fig9} shows the solution to this equation for different initial redshifts (up to $z=4$). As can be seen, the range of modern galaxy baryonic masses ($10^8$ M$_{\sun} - 10^{12}$ M$_{\sun}$) can be achieved by $z=4$ by varying the formation redshift of the progenitor protogalaxies. For the Milky Way ($M_\textnormal{b}\approx10^{11}$ M$_\sun$), this corresponds to a formation redshift of $z_i=20$.

\subsection{Approximate formula for total SMGR}
\label{sec:subsection5.5}

\begin{figure}
\centering
	\includegraphics[width=0.6\columnwidth]{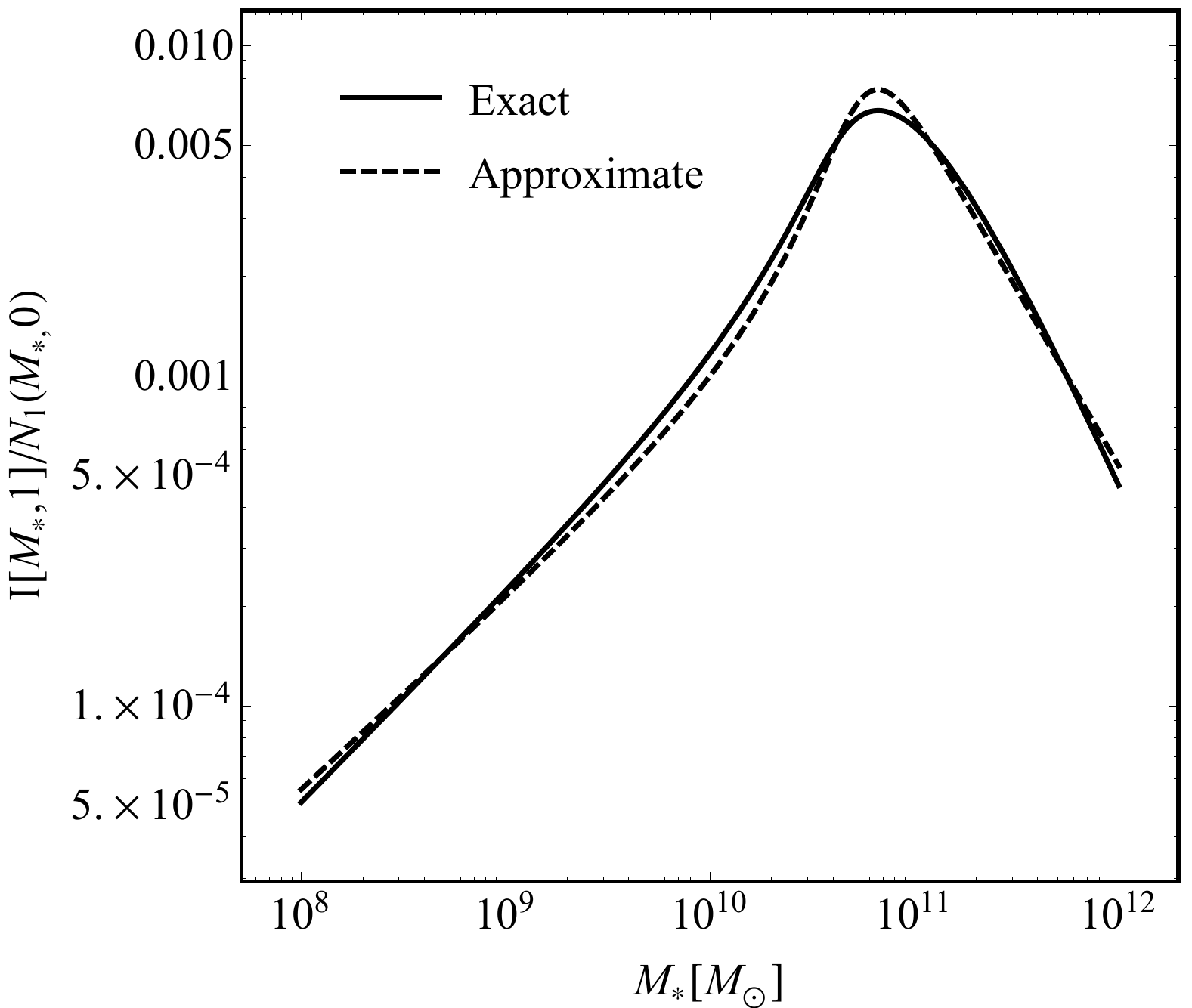}
    \caption{Mass integral divided by the normalisation of the SHMF in the total SMGR as a function of central galaxy stellar mass at redshift zero. The full line was obtained numerically, while the dashed line is the approximate solution of this integral, featured in equations (\ref{eq:ApproximateSMGRI} and \ref{eq:ApproximateSMGRII}).}
    \label{fig:fig10}
\end{figure}

With the mass integral in our two SMGR equations (\ref{eq:SMGRIFinal} and \ref{eq:SMGRIIFinal}) relatively complex, here we find an approximate formula for the total SMGR, with $\mu_\textnormal{max}=\mu(\alpha=1)$ (accounting for all mergers). Alongside this integral, another is featured in the normalisation $N_1$ (equation \ref{eq:NormalisationFinal}). We approximate their appearance in the SMGR with a single function. We find that a double power law with a smooth transition works well. Fig. \ref{fig:fig10} shows both the exact and approximate functions.  Overall, the final formula for the SMGR in model I is
\begin{equation}
   	\frac{\diff M_{*,I}}{\diff t}=\frac{5\pi}{8}pfM\sqrt{G\rho_0}\frac{A(M/M_2)^{\gamma}}{[1+(M/M_2)^\delta]},
	\label{eq:ApproximateSMGRI}
\end{equation}
while for model II we have
\begin{equation}
   	\frac{\diff M_{*,II}}{\diff t}=\frac{5\pi}{4}pfM\sqrt{G\rho_0}\frac{A(M/M_2)^{\gamma}}{[1+(M/M_2)^\delta]}S_0(c,\xi),
	\label{eq:ApproximateSMGRII}
\end{equation}
with the parameters of the approximate mass integral \{$A$, $M_2$, $\gamma$, $\delta$\} equal to \{$1.27\times 10^{-2}$, $10^{12.3}$ M$_{\sun}$, $1.36$, $1.78$\}, respectively.

\subsection{Merger rate}
\label{sec:subsection5.6}

The merger rate can be found in the same manner as the SMGR (equation \ref{eq:SMGRGeneral}). We integrate the individual merger rate, after a multiplication by the mass and spatial distributions. The only sensible choice for the individual merger rate (if we are to count mergers) is the same as in model I for the SMGR$-$the delta function $\delta(t-\tau)$. By a similar procedure as for the SMGR, the  merger rate is then given by
\begin{equation}
   	\frac{\diff N}{\diff t}=\frac{5\pi}{8}pN_1\sqrt{G\rho_0}I_2(M,\alpha),
	\label{eq:MergerRate}
	\end{equation}
\begin{equation}
   	I_2(M,\alpha)=\int_{\mu_\textnormal{min}}^{\mu_\textnormal{max}}\mu^{-0.8}e^{-12.27\mu^3}\ln\bigg(1+\frac{1}{\mu}\bigg)\diff \mu,
	\label{eq:MassInt2}
\end{equation}
where the boundaries $\mu_\textnormal{min}$ and $\mu_\textnormal{max}$ determine the interval of interest. The function
\begin{equation}
   	f(\mu)=\mu^{0.2}e^{-12.27\mu^3}\ln\bigg(1+\frac{1}{\mu}\bigg)
	\label{eq:IntegrandFunc}
\end{equation}
rises very quickly to reach the value $\approx1.85$ (by $\mu=0.01$) and then falls almost linearly to 0 by $\mu=0.55$. We thus choose a linear approximation,
\begin{equation}
   	f(\mu)\approx1.85(1-1.82\mu).
	\label{eq:IntegrandFunc}
\end{equation}
The integrand is equal to $f(\mu)/\mu$. With this approximation, the merger rate can be evaluated analytically to yield:
\begin{equation}
   	\frac{\diff N}{\diff t}=\frac{29}{8}pN_1\sqrt{G\rho_0}\Big[\ln\frac{\mu_\textnormal{max}}{\mu_\textnormal{min}}-1.82(\mu_\textnormal{max}-\mu_\textnormal{min})\Big].
	\label{eq:MergerRateFinal}
\end{equation}
For major mergers, the two integral boundaries are $\mu_\textnormal{min}=\mu(\alpha=0.25)$ and $\mu_\textnormal{max}=\textnormal{min}\{0.55,\mu(\alpha=1)\}$, where the min function ensures that the upper boundary never goes above 0.55, which is the mass ratio at which our approximated integrand (equation \ref{eq:IntegrandFunc}) becomes zero. For larger mass ratios, the integrand is negative and must not be included. For minor mergers, the appropriate choice is $\mu_\textnormal{min}=\mu(\alpha=0.01)$. The upper boundary is given by the usual $\mu_\textnormal{max}=\mu(\alpha=0.25)$.

\begin{figure}
\centering
	\includegraphics[width=0.6\columnwidth]{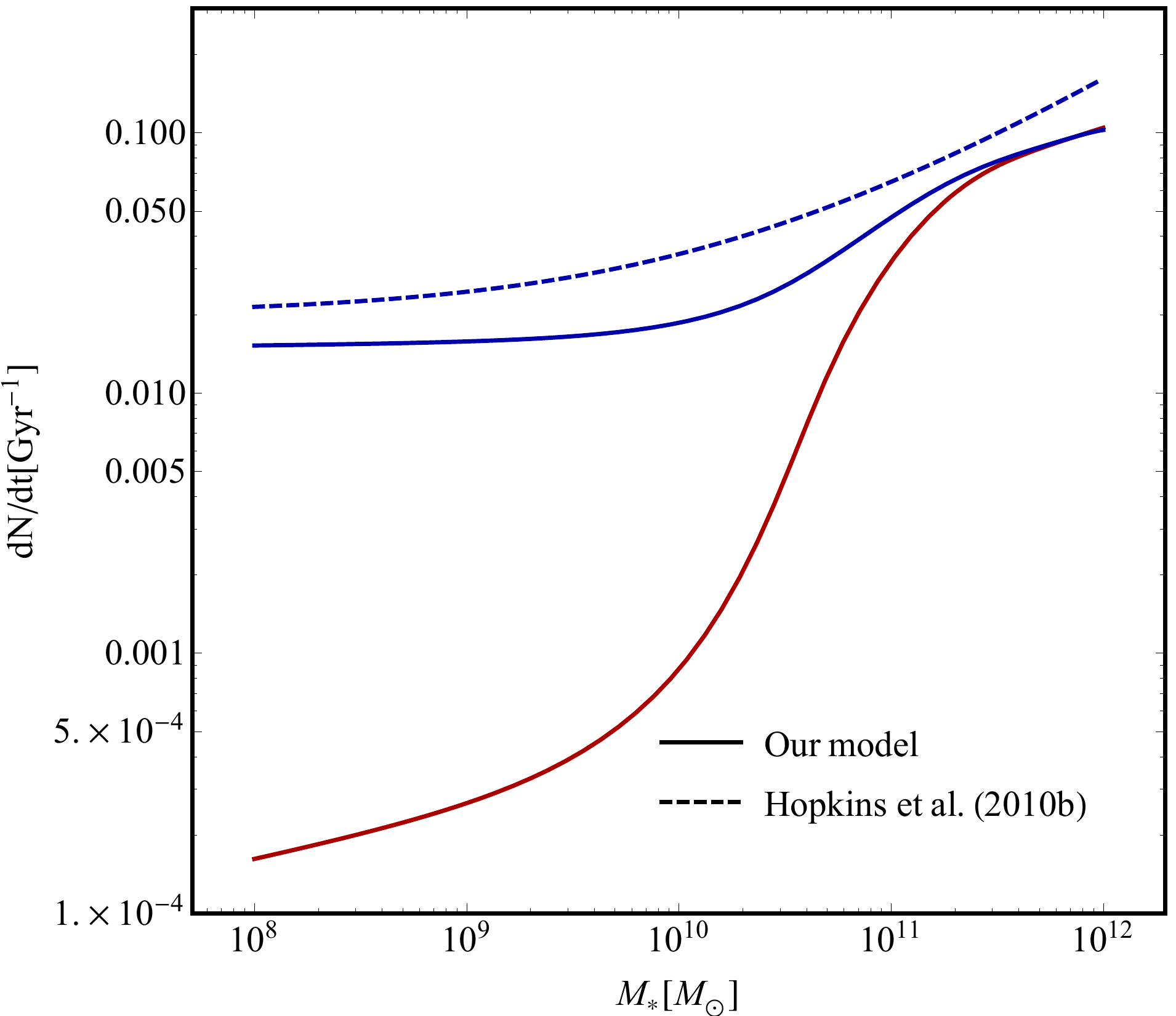}
    \caption{Merger rates of galaxies as functions of stellar mass at redshift $z=0$. Full lines represent our predictions, while the dashed line is the result from empirical studies \protect\cite{Hopkins2010b}. The red line represents our generic merger rate at a given stellar mass (equation \ref{eq:MergerRateFinal}), while the blue line gives the weighted merger rate for galaxies above a given stellar mass (equation \ref{eq:WeightedMergerRate}).}
    \label{fig:fig11}
\end{figure}

Fig. \ref{fig:fig11} shows major merger rates at $z=0$ as a function of stellar mass. The red line gives the major merger rate, obtained as described above. This clearly does not agree with the results of \cite{Hopkins2010b}, where a robust empirical model is presented. However, this is a matter of different definitions$-$the merger rate given in \cite{Hopkins2010b} is defined as the merger rate for galaxies larger than stellar mass $M_*$, while ours is simply the merger rate for galaxies of that stellar mass. In order to make a valid comparison, we need to take a weighted mean of our merger rate over the galaxy population in the interval $M_*-10^{12}$ M$_{\sun}$, where the upper bound of the interval is somewhat arbitrary (chosen to represent the upper end of the galaxy population). This corresponds to computing the following:
\begin{equation}
   	\langle \frac{\diff N}{\diff t}\rangle=\frac{1}{n}\int_{M_*}^{10^{12}\hspace{0.5mm}\textnormal{M}_{\sun}}\frac{\diff N}{\diff t}\frac{\diff n}{\diff M_*'}\diff M_*',
	\label{eq:WeightedMergerRate}
\end{equation}
where $n$ is given by
\begin{equation}
   	n=\int_{M_*}^{10^{12}\hspace{0.5mm}\textnormal{M}_{\sun}}\frac{\diff n}{\diff M_*'}\diff M_*'.
	\label{eq:TotalNumDensity}
\end{equation}
In both integrals, $\diff n/\diff M_*'$ represents the galaxy number density (galaxy mass function). We choose this from \cite{Tomczak2014}. The results of this integration are shown as the blue line on Fig. \ref{fig:fig11}. The rough shape follows that of \cite{Hopkins2010b}, although the normalisations differ by a factor of $\approx$60\%. However, the results in \cite{Hopkins2010b} feature an error of factor 2 across all masses. Our prediction is within this margin of error.

We note that a similar procedure to the one presented in this subsection, as well as in Section \ref{sec:section4} for the SMGR, can be used to calculate other quantities with a simple substitution of the relevant quantity in the integral over mass and spatial distributions. Namely, the dark matter accretion rate (within the optical disk of the galaxy) is given by the integral of $m/\tau$ (or $m\delta(t-\tau)$, depending on the model). This will also approximate the total mass accretion rate well since the mass fraction of gas and stars with respect to dark matter in satellite populations is negligble in this context. 

\subsection{Potential model improvements}
\label{sec:subsection5.7}

While our model does reproduce the current SMGR and merger rate (within the margin error) as found in simulations, as well as the relative contributions of minor and major mergers to the total SMGR, there are various points of potential improvement. Although these are all beyond the scope of this paper, we discuss them here for the sake of completeness. 

\subsubsection{Uncertainties in the model}
\label{sec:subsubsection5.7.1}

While many of the relations used in our work (such as those presented in Section \ref{sec:section2}) come with a well determined uncertainty as a function of stellar (or halo) mass and redshift (e.g. SHMR or SHMF), there are many relations that we use that do not have well determined errors. Furthermore, we make certain assumptions that have intrinsic error which is rather hard to quantify.

An example of this is the assumption that Chandrasekhar's dynamical friction formula (equation \ref{eq:Chandrasekhar}) is applicable to all subhaloes and haloes, whereas there is certainly error associated to this assumption. This error would need to be studied not only as a function of subhalo-to-halo mass ratio $\mu$, but also the environment of the subhalo (such as the DM profile of the host) and the halo mass of the host. There has been some work dedicated to the error in infall time-scale formulas, although simulations used in this studies have generally been restricted to relatively high-mass host haloes \cite{BK2008, Jiang2008}.

With some errors accounted for and others not, we find no point in presenting partially complete predictions with regards to the error in our assumptions. However, below we outline many of the assumptions made that clarify where the various sources of error lie.

\subsubsection{Other points of improvement}
\label{sec:subsubsection5.7.2}

Our model is currently ill-equiped to deal with the transitional redshifts (between $z=0$ and higher redshifts where galaxy formation takes place). This is due to the fact that the probability $p$ of subhaloes hosting galaxies is expected to grow (up to $p=1$) with redshift. In addition, instantaneous merging (model I) is expected to model real mergers better than continuous merging (model II) as redshift increases, but this transition is probably smooth in nature.

A proper study at higher redshift also requires correct modelling of the satellite SHMR (equation \ref{eq:SatSHMR}). We assumed that the factor which accounts for the difference between the satellite and central galaxy SHMRs is constant with redshift, while it is generally expected that the two SHMRs differ less and less as the redshift in consideration approaches the formation redshift of the galaxy (due to tidal stripping having less time to alter the SHMR).

The expected stellar mass in the mass integral is assumed to be proportional to the satellite SHMR for all subhalo masses. However, it can be argued that as the mass ratio $\mu=m/M$ approaches unity, the stellar mass of the `subhalo' should be given by the central SHMR instead of the satellite one. This is equivalent to saying that mergers of similar mass take place between field galaxies. If this is the case, there should in general be a smooth transition from the satellite to the central SHMR centred on some mass ratio $\mu_t$. We have tested various fudicial transitions and found this to only affect the SMGR in the very high mass end ($M_*>10^{11}$ M$_{\sun}$). This is expected, given the fact that this is the regime where the two SHMRs differ most. Curiously, assuming no transition (as we did earlier) produces the best agreement with simulations.

Most of the remaining points of weakness in our model are related to the dynamical friction infall time-scale. The usual choice for the Coulomb logarithm exhibits only a dependence on mass; $\ln \Lambda=\ln (1+M/m)$, while it should technically be allowed to vary with current distance $r$ \cite{Zentner2005, Fujii2006}. Tidal stripping affects this even more by changing the mass of the subhalo, whereas we have assumed it to be constant up until the merger event. While this is a relatively strong assumption, tidal stripping was accounted for in fits done in \cite{Jiang2008}, whose dependence of infall time-scale on orbital angular momentum we used in Section \ref{sec:section3}. The authors of that work used the same expression for the Coulomb logarithm and found good fits to simulation results. 

In Newton's second law for rotational motion (equation \ref{eq:AngMomentum}), we assumed the subhalo moment of inertia $I$ to be given by the point-mass formula $I=mr^2$, while actual subhaloes are extended. This means that the moment of inertia is given by
 \begin{equation}
   	 I=mr^2+I_0(c_\textnormal{sh}),
	\label{eq:MomentOfInertia}
\end{equation}
where $c_{\textnormal{sh}}$ is the concentration of the subhalo. While $I_0$ can be calculated analytically, the calculation of the circular infall time-scale is complicated significantly if it assumed to be non-zero. In general, we expect the inclusion of $I_0(c_\textnormal{sh})$ to be signficant only for high-mass satellites of even higher-mass centrals. This is due to the fact that larger subhaloes are less concentrated, which means their moment of inertia is larger.

We assumed that orbital circularity $\epsilon$ is constant across all subhaloes, while it is in reality a relatively complex function of the mass ratio $\mu$ \cite{Taffoni2003}. We did not use this form of $\epsilon$ in order to avoid nested numerical integrations$-$the integral over the probability distribution $p(\epsilon)$ (equation \ref{eq:DistrEpsilon}) cannot be seperated from the mass integral in the SMGR, if $\epsilon$ is a function of $\mu$.

Finally, the probability $p$ of a subhalo hosting any satellite plays the role of a free multiplicative parameter in our model. While this can be considered a flaw, our model does offer an independent way of finding this quantity through fitting our SMGR to be the same as that given by simulations, as we have done. It should be noted, however, that the value of $p$ found in this way is dependent on how non-circularity is accounted for in the calculation of the average infall time-scale for a given subhalo (as discussed in Section \ref{sec:subsection3.2}). Specifically, different dependences of the infall time-scale on the circularity parameter $\epsilon$ will lead to different average infall time-scales, with these time-scales differing only by a multiplicative factor (due to an integration over the probability distribution of $\epsilon$). This factor is then carried over to the SMGR through its definition (equation \ref{eq:SMGRGeneral}), which then affects our choice of $p$, since it is another multiplicative factor itself.

\section{Summary and conclusions}
\label{sec:section6}

In this work, we studied the infall of a rigid subhalo in an NFW dark matter profile of a host halo. Using Chandrasekhar's formula for dynamical friction, in addition to previous results on the dispersion of DM particles, we derived the infall time-scale for a subhalo (the mass of which is assumed to be constant) on a circular orbit. Our results are in agreement with previous ones, although they are more general in that they account for cases of initial radius that differ from the host halo radius. While the derived infall time-scale is unrealistic due to the assumptions made, we expect the generalised infall time-scale (equation \ref{eq:FinalTimescale}) to yield correct values since we used formulas for circularity that were obtained through simulated studies of realistic subhaloes (non-rigid and subject to tidal stripping). 

Using this infall time-scale in unison with the expected subhalo mass function, and under the assumption that satellites follow an NFW profile, we derived the expected stellar mass growth rate due to all merger ratios smaller than $\alpha$, as a function of central galaxy stellar mass $M_*$. We did this for two different models, one where galaxies merge instantaneously (given by equation \ref{eq:SMGRIFinal}), and one where they merge continuously (given by equation \ref{eq:SMGRIIFinal}). We also derived galaxy merger rates in a similar fashion. Using the obtained formulas and the subsequent analysis, we conclude the following:
\begin{itemize}
\item The model II (continuous merging) total SMGR follows the results from the Millennium-II simulation closely, predicting little growth in the low-mass range ($10^8$ M$_{\sun} - 10^{10}$ M$_{\sun}$) and increasingly larger growth in the high-mass range ($10^{10}$ M$_{\sun} - 10^{12}$ M$_{\sun}$). Model I (instantaneous merging) agrees with the Millennium-II simulation in the $10^{10}$ M$_{\sun} - 10^{11}$ M$_{\sun}$ range, meaning that galaxies in this range could in principle grow through both kind of processes (violent merging and smooth accretion). Galaxies outside this mass range are constrained to grow through smooth accretion from their satellites.

\item Early-type and late-type galaxies grow differently, with low-mass early-types growing faster than late-types, and high-mass late-types growing faster than early-types. We suspect this to be an additional factor contributing to the relative rarity of high-mass late-types.

\item Minor mergers are the dominant process for a wide range of masses and redshifts, contributing 80$-$90\% of all mass growth for galaxies in the $10^{8}$ M$_{\sun} - 10^{10}$ M$_{\sun}$ range from redshift $z=3$ to $z=0$. For higher masses (>$10^{11}$ M$_{\sun}$), minor mergers contribute 10$-$20\% of mass growth. Overall, minor mergers are expected to have contributed the majority of accreted mass for most galaxies but the most massive ones.

\item Major mergers contribute somewhat to mass growth at all masses. However, they only become dominant for galaxies more massive than $3-4\times10^{10}$ M$_{\sun}$.

\item Our model is consistent with a picture of hierarchal galaxy formation in which all modern galaxies are descended from initial protogalaxies having a mass of $\approx$ $10^6$ M$_\sun$. These protogalaxies started forming from collapsing clouds of baryons after recombination, and they only differ in the redshift at which they formed. Galaxies that formed earlier are thus more massive.

\item Merger rates are well described by equation (\ref{eq:MergerRate}) or its approximation (equation \ref{eq:MergerRateFinal}). Extrapolating our formula to higher redshifts we find that merger rates evolve as $(1+z)^4$.

\item The probability of subhaloes hosting satellites is $p=10\%$. This value is required in order for our model to reproduce the SMGR as found in simulations, and is in good agreement with previous studies motivated by the `missing satellites' problem. 
\end{itemize}
With many previous results from simulations and observations reproduced by our model, both qualitatively and quantitatively, we conclude that dynamical friction is the dominant cause of both minor and major mergers. We tentatively conclude that our model can be used to describe galaxy formation and mass evolution due to mergers.

\acknowledgments

I thank Vernesa Smolčić and Vibor Jelić for their support of this work, as well as Krešimir Tisanić for helpful comments.


\end{document}